\documentclass[iop]{emulateapj}

\usepackage{amsmath}
\usepackage{natbib}
\usepackage{hyperref}
\usepackage{mhchem}
\usepackage{graphicx}
\usepackage{chngcntr}
\usepackage{rotating}
\usepackage{threeparttable}
\newcommand{\htoj}{$\mathrm{H}_2\mathrm{O} - \mathrm{J}$}

\newcommand{\myemail}{gbruno@stsci.edu}
\def\gtsima{$\; \buildrel > \over \sim \;$}
\def\ltsima{$\; \buildrel < \over \sim \;$}
\def\gtrsim{\lower.5ex\hbox{\gtsima}}
\def\lesssim{\lower.5ex\hbox{\ltsima}}
\def\kr{$R_\mathrm{p}/R_\star$}

\slugcomment{}

\shorttitle{Starspot occultations in infrared transit spectroscopy}
\shortauthors{Bruno et al.}

\begin{document}


\title{Starspot occultations in infrared transit spectroscopy: the case of WASP-52\lowercase{b} }


\author{Giovanni Bruno\altaffilmark{1}, Nikole K. Lewis\altaffilmark{1,2}, Kevin B. Stevenson\altaffilmark{1}, Joseph Filippazzo\altaffilmark{1}, Matthew Hill\altaffilmark{1}, Jonathan D. Fraine\altaffilmark{1}, Hannah R. Wakeford\altaffilmark{1}, Drake Deming\altaffilmark{3}, Mercedes L\'{o}pez-Morales\altaffilmark{4}, Munazza K. Alam\altaffilmark{4}}
\affil{$^1$ Space Telescope Science Institute, 3700 San Martin Drive, Baltimore, MD 21218, USA}
\affil{$^2$ Department of Earth and Planetary Sciences, Johns Hopkins University, Baltimore, MD, USA}
\affil{$^3$ Department of Astronomy, University of Maryland at College Park, College Park, MD 20742, USA}
\affil{$^4$ Harvard-Smithsonian Center for Astrophysics, 60 Garden Street, Cambridge, MA 01238, USA}


\altaffiltext{1}{\myemail}


\begin{abstract}
Stellar activity is one of the main obstacles to high-precision exoplanet observations and has motivated extensive studies in detection and characterization problems. Most efforts focused on unocculted starspots in optical transit spectrophotometry, while the impact of starspot crossings is assumed to be negligible in the near-infrared. Here, we present \textit{HST}/WFC3 transit observations of the active star WASP-52, hosting an inflated hot Jupiter, which present a possible starspot occultation signal. By using this data set as a benchmark, we investigated whether the masking of the transit profile distortion or modeling it with both a starspot model and a Gaussian process affects the shape of the transmission spectrum. Different methods produced spectra with the same shape and a robust detection of water vapor, and with $\lesssim 1 \sigma$ different reference radii for the planet. The solutions of all methods are in agreement and reached a similar level of precision. Our WFC3 light curve of WASP-52b hints that starspot crossings might become more problematic with \textit{JWST}'s higher sensitivity and complete coverage of the transit profile.
\end{abstract}





\section{Introduction}\label{intro}

Transiting extrasolar planets offer the unique opportunity to study stellar activity via the detection of starspots on the surface of their hosts \citep{silva2003}. Since the \textit{Hubble Space Telescope} (\textit{HST}) observations of starspots on HD209458 \citep{brown2001,deeg2001}, the \textit{CoRoT} \citep{auvergne2009} and \textit{Kepler} \citep{borucki2010} space telescopes opened new possibilities in the study both of stellar activity and of star-planet interactions \citep[e.g.][and references therein]{affer2012,mcquillan2014,lanza2014}. These observations presented a variety of cases in which the stellar signal also hampers the interpretation of the exoplanet signal. The transit observations of the extensively studied HD189733 \citep{pont2007}, CoRoT-2 \citep{alonso2008}, CoRoT-7 \citep{leger2009} and Kepler-17 \citep{desert2011} showed how unocculted and occulted activity features deform the shape of a transit by modulating the the host star's brightness.

Unocculted starspots (i.e., those lying out of the transit chord) can affect the transit baseline and possibly induce an overestimate of the planet radius \citep[e.g.][]{czesla2009}. Occulted spots can affect the transit shape, which can possibly cause an underestimate or an overestimate of the transit depth, depending on the presence of dark or bright spots, respectively \citep[e.g.][]{silva-valio2010,desert2011,bruno2016}. The measurement of other parameters can also be biased, such as the orbital inclination, stellar density and limb darkening coefficients \citep[e.g.][]{leger2009,csizmadia2013}. Published strategies for the correction of the starspot signal in broadband transit photometry span from modeling the stellar surface in a grid \citep[e.g.][]{huber2010}, to maximum entropy regularization \citep[e.g.][and references therein]{lanza2009}, combined planet-starspot modeling \cite[e.g][]{tregloan-reed2015,bruno2016} and Gaussian processes on unocculted starspots \citep[e.g.][]{haywood2014,aigrain2016}. The study of the effect of starspots on the measure of planetary masses through radial velocity is another active area of investigation \citep[e.g.][]{saar1997,hatzes1999,queloz2001,melo2007,desort2007,boisse2011,boisse2012,robertson2014,dumusque2017,feng2017}. 

In transmission spectroscopy, the transit depth of a planet with an atmosphere can be observed to vary with wavelength. This is due to the variation of the atmospheric opacity with wavelength, which in turn affects the radius at which the atmosphere becomes optically thick \citep{seager2000,brown2001,hubbard2001}. During primary transit, unocculted spots are known to mimic the Rayleigh scattering feature at visible wavelengths \citep[][]{pont2013,mccullough2014,oshagh2014,rackham2017}. As a result, consecutive observations of the same targets often yield different results \citep[e.g.][and references therein]{mackebrandt2017}. Here too, various techniques have been adopted to correct for stellar activity, including the modeling of starspot brightness with stellar models \citep[e.g.][]{mccullough2012,sing2016}, transit-spot modeling \citep[e.g][]{mancini2014} and Gaussian processes \citep[e.g.][and references therein]{gibson2013gemini,louden2017,sedhaghati2017}. Recent studies attempted to forecast the biases and uncertainties that unocculted starspots will produce on the \textit{James Webb Space Telescope} (\textit{JWST}) spectra \citep{barstow2015,barstow2015err,zellem2017,serrano2017,deming2017}. 

As they are less prominent at longer wavelengths, occulted starspots are generally considered less problematic in the near-infrared than in the visible. Furthermore, only very cool ($\lesssim 3000$ K) starspots in solar-type stars could contaminate the water absorption feature of a planetary spectrum \citep{pont2013,fraine2014}. It is usually assumed that the portions of a transit profile affected by starspot crossings can be safely removed from the analysis, especially with \textit{HST} data. As this observatory is periodically occulted by the Earth, it only gives a partial phase coverage of transit profiles. Hence, parameters such as limb darkening coefficients and orbital inclination (which in turn affect the measured transit depth) are poorly constrained by observations and are often fixed in the transit profile fitting. \textit{JWST} observations, which will not be limited in phase coverage, will be sensitive to the effect of stellar activity on transit parameters other than the transit depth.

As the cases of e.g. CoRoT-2 \citep{silva-valio2010} and Kepler-117 \citep{desert2011} showed, a large fraction of the transit profile can be affected by starspot occulations, and the variation of these features in time could cause erroneous detections of ``weather'' variability on the planets \citep{pont2013}. Additionally, cool, low-mass stars, which are the main focus of upcoming exoplanet searches such as the \textit{Transiting Exoplanet Survey Satellite} \citep[\textit{TESS},][]{ricker2014}, are mainly convective and often have a larger starspot coverage than solar type stars \citep{chugainov1966,chugainov1971,berdyugina2011,mandal2017}. If several occultations of small -- but still detectable -- starspots affect a single transit profile, masking them will not be a satisfactory strategy. Even if longer wavelengths are less affected by occulted starspots, very cold activity features will be cause of additional concern in the interpretation of planetary spectra.

\textit{HST} observations of active planet host stars are a great opportunity to explore this problem, as in the case of the K star WASP-52. \cite{hebrard2013_w52} observed long-term modulations in photometry and CaII H\&K lines chromospheric emission peaks, suggesting the presence of starspots on the stellar surface. Moreover, gyrochronology and lithium abundance yielded contrasting estimates of the stellar age -- from less than one to several Gyr. Different age indicators might be indicative of planet-induced magnetic and tidal spinning up of the star, with consequences on its overall activity \citep[e.g.][]{shkonlnik2005,dawson2014_tidal,damiani2015}. Alternatively, or simultaneously, enhanced lithium depletion in the star might have happened because of planet-induced alterations of the stellar surface convective mixing \citep{israelian2009,sousa2010}.

The inflated hot Jupiter WASP-52b orbits its star with a 1.7-day period. \cite{hebrard2013_w52} measured a stellar spin-planetary orbit misalignment of $24^{+17}_{-9}$ deg via the Rossiter-McLaughlin effect \citep{rossiter1924,mclaughlin1924}, while \cite{mancini2017}, assuming that their multiple transit observations were affected by the same starspot, calculated a negligible misalignment. The system parameters, summarized in Table \ref{parameters}, make WASP-52b particularly interesting for transmission spectroscopy, as it shows a 2.7\% transit depth in WASP photometry and an estimated scale height of 730 km. This translates in a predicted 440 ppm difference in transit depth corresponding to one atmospheric scale height, at least three times stronger than that of HD189733b \citep{kirk2016}. This motivated ground-based follow-ups by various teams, which led to the likely detection of crossing events of both starspot and bright regions in the visible \citep{kirk2016,mancini2017}. 

Primary transit spectroscopy of WASP-52b in the 0.4-0.9 $\mu$m spectral window resulted in a flat transmission spectrum (i.e., no Rayleigh scattering detection), likely due to a gray cloud which could balance the expected short wavelength slope \citep{louden2017}. The possible contribution of bright spots in the visible spectrum, which could compensate for the scattering slope, should also not be excluded \citep{rackham2017}. In addition, a narrow NaI absorption feature \citep{kirk2016,louden2017,chen2017}, KI absorption and indications of a thermal inversion in the high atmosphere \citep{chen2017} have been found.

We collected \textit{HST}/Wide Field Camera 3 (WFC3) G141 observations of WASP-52 as part of a large program for the analysis of exoplanet atmospheres with \textit{HST} (GO 14260, PI Deming). In this work, we compared the main approaches currently used in the literature to deal with starspot occultations, which likely affect our spectroscopic observations. Our goal was to identify, if any, differences in the transmission spectrum which can be attributed to the choice of the starspot correction method. The data set is presented in Section \ref{observations} and the various models on the band-integrated transit in Section \ref{transitmodeling}. The derivation of the transmission spectrum is discussed in Section \ref{spectrotr}, and the implications of our analysis are presented in the concluding Section \ref{discussion}.

\begin{table}[htb]
\begin{center}
\caption{WASP-52 system parameters, from \cite{hebrard2013_w52}.}
\label{parameters}
\begin{tabular}{ll}
\hline \hline
Spectral type                                    & K2V \\
Stellar $V$ magnitude                            & 12.0 \\
Stellar $J$ magnitude$^{(\ast)}$                            & 10.6 \\
Stellar $H$ magnitude$^{(\ast)}$                            & 10.1 \\
Stellar $K$ magnitude$^{(\ast)}$                            & 10.2 \\
Stellar effective temperature $T_\mathrm{eff, \star}$ [K] & $5000\pm100$ \\
Stellar $\log g$ [cgs]                           & $4.5 \pm 0.1$ \\
Stellar [Fe/H] [dex]                             & $0.03 \pm 0.12$\\
Stellar radius [$R_\odot$]                       & $0.79 \pm 0.02$\\
Stellar density [$\rho_\odot$]          &  $1.76 \pm 0.08$ \\
Stellar rotation period [days]                   & $11.8 \pm 3.3$\\
$\log R'_{HK}$                                   & $-4.4\pm0.2$ \\
Orbital period [days]                            & $1.7497798\pm0.0000012$ \\
Transit duration [days]                          & $0.0754\pm0.0005$ \\
Radius ratio $R_\mathrm{p}/R_\star$             & $0.1646\pm0.0012$ \\
Orbital inclination [$^{\circ}$]                    & $85.35\pm0.20$\\
Scaled semi-major axis $a/R_\star$               & $7.3801\pm0.0148$ \\
Orbital eccentricity                             & 0 (fixed)  \\
Spin-orbit misalignment [$^\circ$]               & $24^{+17}_{-9}$\\
Planet mass [$M_J$]                              & $0.46\pm0.02$   \\
Planet radius [$R_J$]                            & $1.27\pm0.03$   \\
Planet density [$\rho_J$]                        & $0.22\pm0.02$   \\
Planetary equilibrium temperature [K]                              & $1315\pm35$     \\ 
Planet surface gravity [$\mathrm{m \, s}^{-2}$]  & $6.46\pm0.45$ \\
\hline
\end{tabular}
\end{center}
\begin{tablenotes}\footnotesize
    \item $^{(\ast)}$ From the CMC15: Carlsberg Meridian Catalogue (\texttt{http://svo2.cab.inta-csic.es/vocats/cmc15/}, retrieved through \textit{VizieR} on the Strasbourg astronomical Data Center (\texttt{http://cds.u-strasbg.fr/}).
\end{tablenotes}
\end{table}

\section{Observations and data reduction}\label{observations}


The WFC3 G141 spectra of WASP-52 are publicly available on the Mikulski Archive for Space Telescopes\footnote{\dataset[10.17909/T93D5K]{https://doi.org/10.17909/T93D5K}.} and cover a single visit of WASP-52b, obtained on 2016 August 28. The transit duration is $\sim 1.8$ hr, which required us four {\em HST} orbits to encompass the necessary out-of-transit phase.

The target was acquired in forward scanning direction only, with a scan rate of 0.035 arcsec s$^{-1}$ (0.27 pixel s$^{-1}$). We used the frames in the IMA format, where each file contains eight 22.3 s long non-destructive reads (NDR), for a total exposure time of about 134 s. The $256 \times 256$ pixels aperture was used in SPARS25 mode, yielding $\sim 30,000$ average electron counts per exposure. A first calibration of the raw images and correction for instrumental effects was carried out by the \texttt{calwf3} pipeline, version 3.3.

\begin{figure}[htb]
\epsscale{1.3}
\plotone{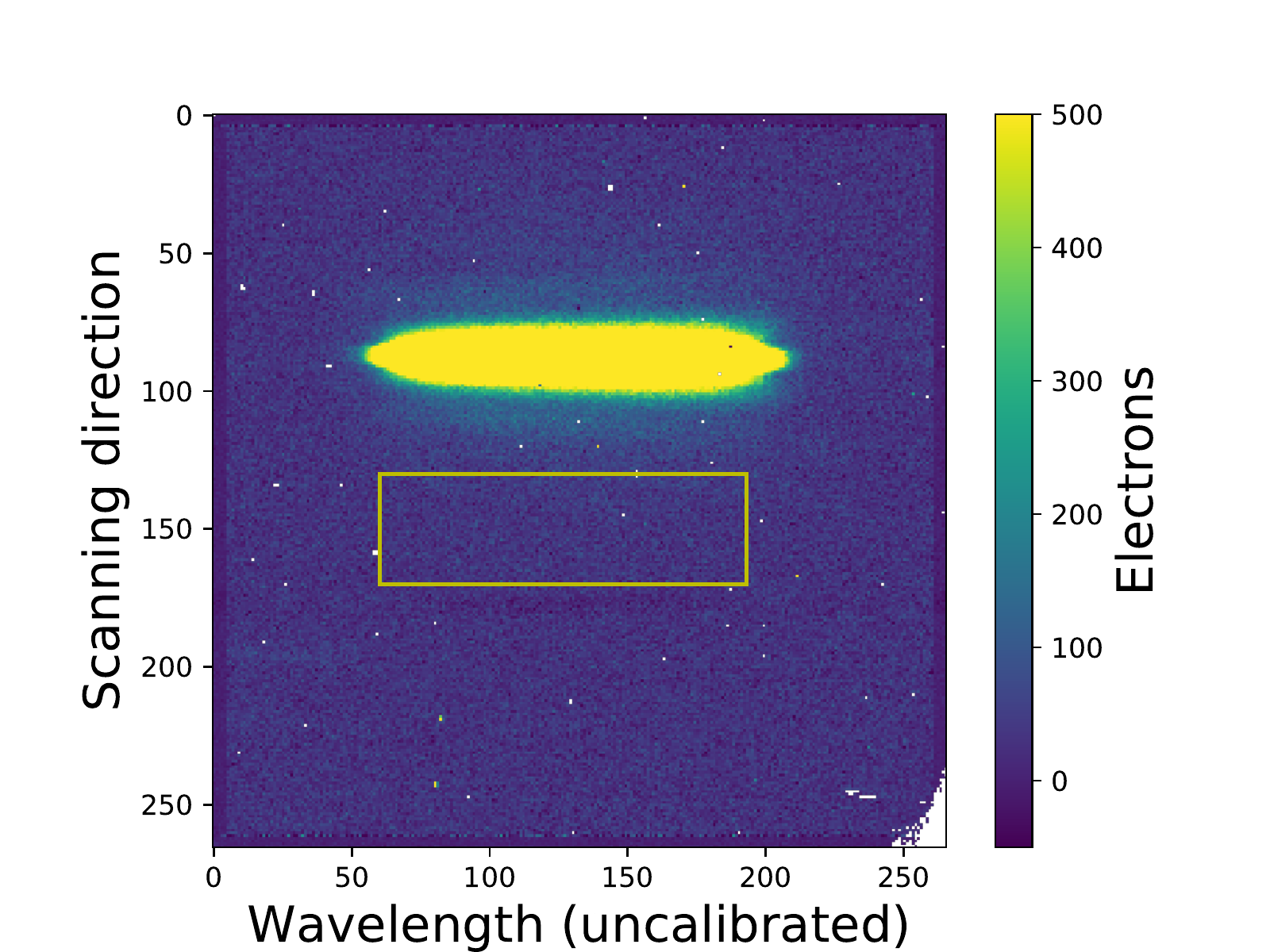}
\caption{Background flux in a WASP-52 frame. The $x$-axis shows the values of column pixel, prior to wavelength calibration. The yellow box indicates the region used for background estimation. The plotted values are limited in the range indicated in the right column.}
\label{bkg1}
\end{figure}

\begin{figure}[htb]
\epsscale{1.15}
\plotone{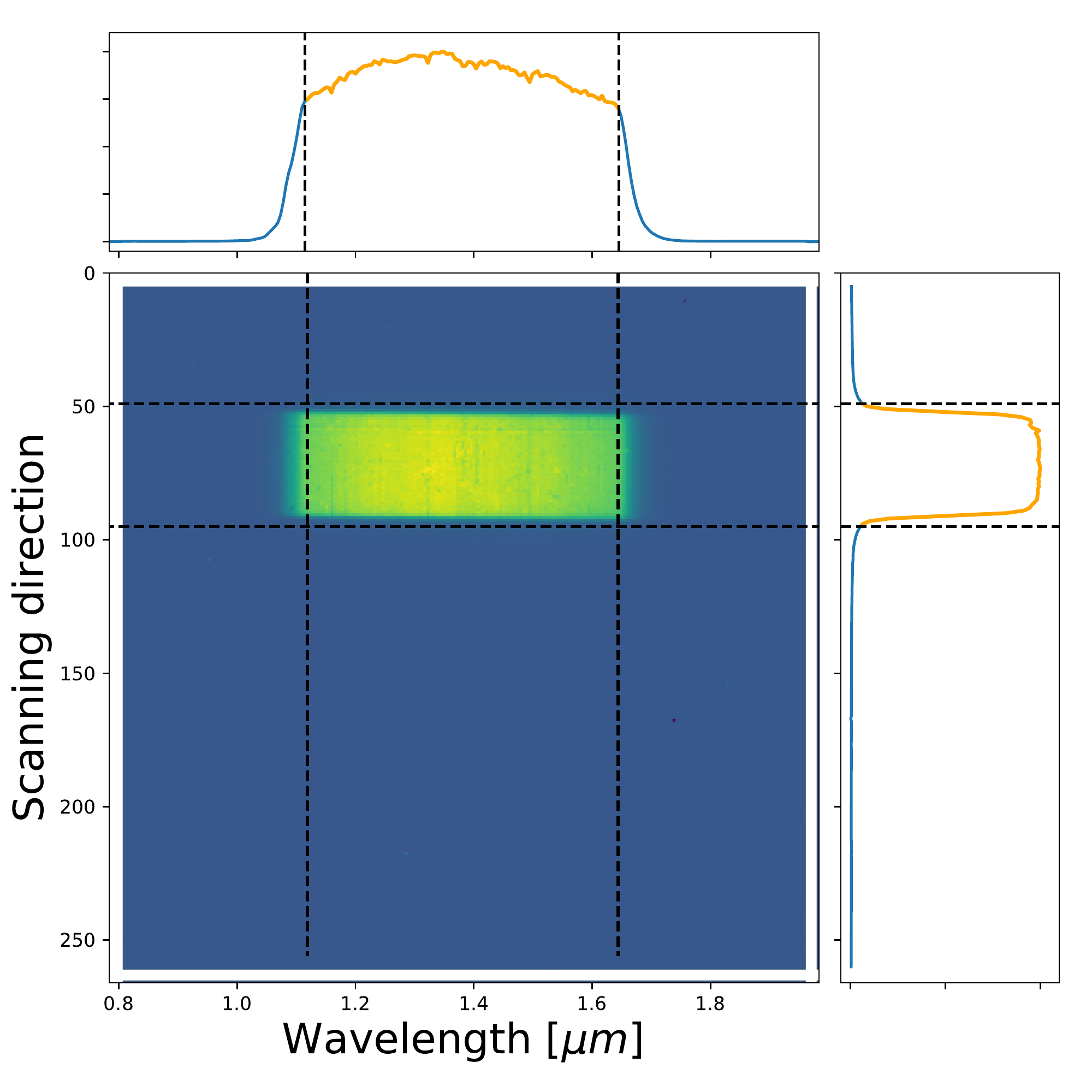}
\caption{The 2D spectrum presented in Figure \ref{bkg1} after background subtraction and optimal extraction. On the top plot, the 1D spectrum after integration over the frame columns. On the right, the spectral trace in the cross-dispersion direction. The dashed lines and the orange color show the limits chosen for wavelength and aperture integration. The flux range goes from 0 to about 35000 electrons.}
\label{reducedsp}
\end{figure}

Wavelength calibration, operations on the NDR, background subtraction, cosmic ray and bad pixels rejection, and inspection for drifts were carried out following standard procedures as described in \cite{bruno2018_clouds} and references therein. In particular, the background subtraction was performed column-by-column, by selecting a region of the detector placed below the stellar spectrum, as shown in Figure \ref{bkg1}. Figure \ref{reducedsp} shows the spectrum after the reduction of one of the IMA files.

\begin{figure}[htb]
\epsscale{1.25}
\plotone{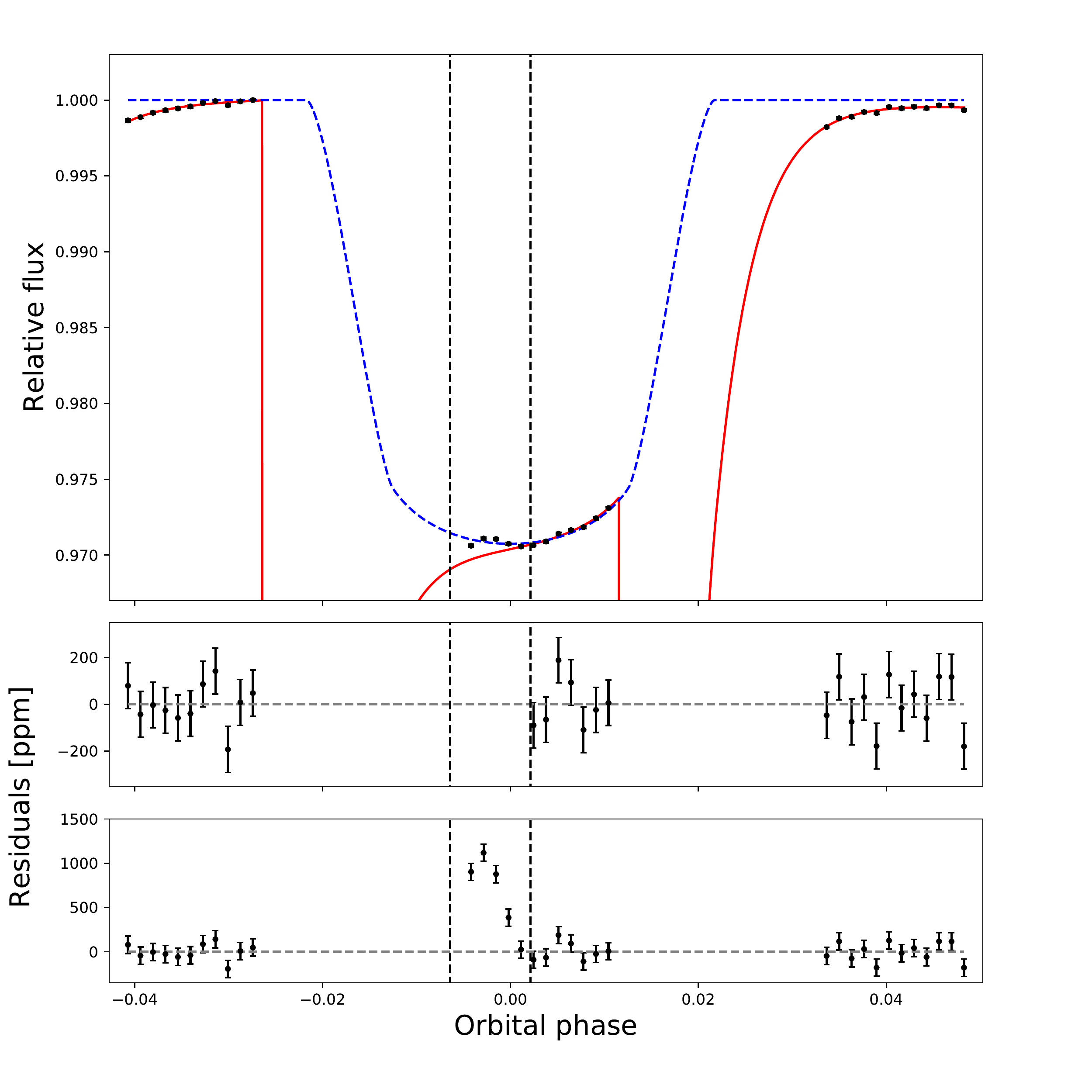}
\caption{Band-integrated light curve of WASP-52b, best-fit starspot-free model including transit and systematics (red line, discussed in Section \ref{whitetr}), and the same model without systematics (dashed blue line). The transit profile distortion excluded from the fit is included within vertical dashed lines. In the two lower panels, different scales are used to show the residuals on the masked and the complete data set.}
\label{white}
\end{figure}

The integrated flux between 1.115 and 1.645 $\mu$m, in orange in the top panel of Figure \ref{reducedsp}, yielded the photometric flux measurement corresponding to each point of the band-integrated, or ``white'', light curve. The resulting white light curve is shown in Figure \ref{white}, where a possible starspot occulation is visible at about orbital phase 0 the models presented in Figure 3 are discussed in section \ref{whitetr}).\footnote{The raw light curve is available in tabular format as Data behind the Figure (DbF).} As such feature is localized, we can exclude that it is part of the periodic \textit{HST} breathing, exponential ramp and visit-long slopes reported by many groups \citep[e.g.][and references therein]{wakeford2016}. Multiple reductions of the IMA files, where the background regions and the rejection threshold for the cosmic rays and hot pixels were varied, produced the same feature.

Spectroscopic light curves were obtained by integrating the stellar spectra in four to ten pixels-wide channels. Repeating the transmission spectrum extraction for various bin sizes allowed us to determine the robustness of the reduction. While the bluemost wavelength used for the integration was kept fixed, different binnings resulted in the redmost wavelength varying by $\sim2$ pixels ($\sim 10$ nm). The profiles resulting from different binnings are all very similar in absolute and relative values. We finally adopted a bin size of 6 pixels, but the quality of the white light curve fit or the shape of the transmission spectrum were not affected by the choice of the bin size.




\section{Band-integrated transits}\label{transitmodeling}

\subsection{Masking the starspot feature}\label{whitetr}

We followed the standard practice \citep[e.g.][]{deming2013} of discarding the first {\em HST} orbit from each transit, which is affected by considerably different systematics than the remaining ones. For the same reason, we also rejected the first data point of every orbit.

An initial attempt to include the starspot-like distortion in a standard model including both the transit profile and the instrument systematics revealed that such ``bump'' prevents to achieve a good fit of the exponential ramp (reduced chi square $\tilde{\chi}^2 \simeq 2.7$). We repeated the optimization after masking the corresponding data points, obtaining a much better reduced chi square ($\tilde{\chi}^2 = 1.66$). We adopted a systematics model including a visit-long second order polynomial and an exponential ramp \citep[e.g.][]{stevenson2014_gj},
\begin{equation}
    S(t) = C(1 + r_0\theta + r_1\theta^2)(1 - e^{r_2\phi + r_3} + r_4\phi),
\label{systmodel}
\end{equation}
where we fitted for $C$ and $r_{0-4}$, and $\theta$ and $\phi$ represent the planetary and \textit{HST} phase, respectively.\footnote{It was necessary to add a shift to the \textit{HST} orbital phase, $\phi = 2\pi [(t - 0.025\mathrm{d}) \mod P_{HST}]/P_{HST}$.} We remark that the normalization constant $C$ includes our ignorance of the stellar brightness at the moment of observation compared to the ``quiet'' stellar surface. Long-time photometric monitoring of the star would have given some constraint on this parameter and allowed us to use \cite{czesla2009}'s or \cite{huitson2013}'s prescription to reduce the risk of overestimating the transit depth because of unocculted starspots. Due to this lack of information, we simply normalized the transit to its maximum value before fitting for $C$.

The transit profile was parametrized with a \cite{mandelagol2002} model, implemented in the \texttt{PyTransit} software \citep{parviainen2015}. We fitted for the planet-to-star radius ratio \kr~ and the transit midpoint $t_0$, fixed the orbital period $P$ to \cite{hebrard2013_w52}'s mean results, and fixed the eccentricity $e$ to 0 as in the discovery paper. We set the scaled orbital semimajor axis $a/R_\ast$ and orbital inclination $i$ as jump parameters, and used \citeauthor{hebrard2013_w52}'s results as priors. Even if data does not cover the transit edges, in this way we determined the dependency of our results on parameters which are little constrained by observations. 

As the limb darkening (LD) coefficients have an important effect on \kr~ and can deviate from model predictions in the case of spotted stars \citep{csizmadia2013,espinoza2015}, they were also left free to vary under the control of priors. We used a quadratic LD law \citep[e.g.][and references therein]{howarth2011} and the \texttt{ExoCTK} package\footnote{\texttt{https://exoctk.stsci.edu/}} to linearly interpolate the LD coefficients from PHOENIX specific intensity stellar models \citep{husser2013} for the $1.115-1.645 \, \mu$m wavelength range. We adopted the mean values of $T_\mathrm{eff}, \, \log g$ and $[\mathrm{Fe/H]}$ from \cite{hebrard2013} and scaled the stellar intensities to the G141 bandpass. For most channels, the \texttt{ExoCTK} LD package provided uncertainties on the coefficients of the order of $10^{-2}$ or less, but we used Gaussian priors with the model-derived values as means and 0.1 as standard deviation.

Our priors are summarized in Table \ref{priortransits}. For this and all the following fits, we started the exploration of the parameter space by performing a least-squares minimization using a L-BFGS-B optimization algorithm \citep{byrd1995,zhu1997,scipy} and proceeded with Markov chain Monte Carlo (MCMC) sampling of the posterior distributions with \texttt{emcee} \citep{goodman-weare2010,foreman-mackey2013}. Here and in the following fits, 200 walkers were used in 2000 iterations-long chains, with 500 previous iterations as burn-in. All chains were then thinned by a factor 10 and merged in a single chain. As suggested by \cite{foreman-mackey2013}, the well-mixing of the chains was inspected by measuring the autocorrelation time of the merged chain and ensuring that such chain was at least 50 times longer than its autocorrelation time for each free parameter. With the number of walkers and iterations we used, the chains were always from hundreds to thousands autocorrelation times-long.

\begin{table*}[htb]
\begin{center}
\caption{Priors for MCMC with the spot-free, spot-transit, and GP models. $\mathcal{U}(a, b)$ denotes a Uniform distribution between $a$ and $b$, $\mathcal{G}(a, b)$ indicates a Gaussian prior with mean value $a$ and standard deviation $b$ and $\Gamma(a, b)$ denotes a Gamma distribution with shape parameter $a$ and scale parameter $b$.}
\label{priortransits}
\begin{tabular}{llll}
\hline
\hline
Parameter                                       & Spot-free transit model   & Spot-transit model  & GP \\
\hline
Radius ratio \kr                             & $\mathcal{G}(0.165, 0.01)$ & $\mathcal{G}(0.165, 0.01)$ &  $\mathcal{G}(0.165, 0.01)$ \\
Transit midpoint $t_0$ [BJD$_\mathrm{UTC} - 2457629.13125$]     & $\mathcal{G}(0.072, 0.01)$ & -$^{(a)}$ & $\mathcal{G}(0.072, 0.01)$\\
Planet initial mean anomaly $M$ [$^\circ$]              & - & $\mathcal{U}(230, 270)$  & -\\
Scaled semi-major axis $a/R_\ast$             & $\mathcal{G}(7.38, 0.11)$ & - $^{(b)}$ & $\mathcal{G}(7.38, 0.11)$ \\
Stellar density $\rho_\ast$ [$\rho_\odot$]                  & - & $\mathcal{G}(1.76, 0.08)$ & -  \\
Orbital inclination [$^\circ$]                  & $\mathcal{G}(85.35, 0.20)$ &  $\mathcal{G}(85.35, 0.20)$ &  $\mathcal{G}(85.35, 0.20)$\\
Orbital eccentricity (fixed)                        & 0 & 0 & 0 \\
Orbital period [days] (fixed)                   & 1.7497798 & 1.7497798 & 1.7497798\\
Linear LD coefficient $u_a$        & $\mathcal{G}(0.252, 0.1)$ & $\mathcal{G}(0.252, 0.1)$ & $\mathcal{G}(0.252, 0.1)$\\
Quadratic LD coefficient $u_b$      & $\mathcal{G}(0.197, 0.1)$ & $\mathcal{G}(0.197, 0.1)$ & $\mathcal{G}(0.197, 0.1)$ \\
Stellar inclination  [deg] (fixed)				& - & 61.35$^{(c)}$ & - \\
Spot brightness contrast $f_c$                              & -  &  $\mathcal{U}(0, 1)$ & - \\
Spot longitude $\lambda$ [$^\circ$]               & - & $\mathcal{G}(270, 30)$  & -\\
Spot latitude $\phi$ [$^\circ$]                   & - & $\mathcal{G}(50, 20)$    & - \\
Spot angular size $\alpha$ [$^\circ$]                     & - & $\mathcal{U}(0, 20)$ & - \\
Spot contrast $f_c$                                     & - & $\mathcal{U}(0, 1)$$^{(e)}$ & - \\
Systematics $r_0, r_3, r_4$                               & $\mathcal{U}(-\infty, \infty)$ & - & - \\
Systematics $r_1, r_2$                                     & $\mathcal{U}(-\infty, 0)$ & - & - \\
Normalization $C$                                          & $\mathcal{U}(0.5, 1.5)$ & $\mathcal{U}(0.5, 1.5)$ & - \\
Inverse characteristic length scale $\eta$              & -  & - & $\Gamma (1, 200)$\\
Period $P$     					      & - & - & $\Gamma (1, 10^{-3})$\\
\hline
\end{tabular}
\end{center}
\begin{tablenotes}\footnotesize
    \item $^{(a)}$ The \texttt{KSint} model measures the mean anomaly instead of the transit midpoint.
    \item $^{(b)}$ \texttt{KSint} uses $\rho_\ast$ instead of $a/R_\ast$.
    \item $^{(c)}$ Using the planetary orbit inclination and the sky-projected angle between the planetary orbital axis and the stellar rotation axis from \cite{hebrard2013_w52} to determine the starting value.
     \item $^{(e)}$ This is equivalent to excluding scenarios with faculae.
\end{tablenotes}
\end{table*}

The results of this MCMC optimization are reported in Table \ref{whiteres} and the best fit model is presented in Figure \ref{white}. The \kr~ found by this fit is within the $1.5\sigma$ credible intervals of \cite{hebrard2013_w52}, \cite{kirk2016}, \cite{mancini2017} and \cite{louden2017}'s results. Thanks to the space-borne observation, we achieved a four times better precision on the \kr~ uncertainty with respect to the discovery paper. However, because of lack of information on the out-of-transit stellar brightness modulation -- which requires long-baseline photometric observations of the host star --, we have no constraint on possible unocculted starspots in the transit. Given the level of activity of the star, such unocculted starspots are likely to have affected the transit baseline, so that our \kr~ is probably overestimated \citep{czesla2009}.
 
In the following, we discuss two methods to include the transit distortion in the fit. In the first, we assumed the distortion to be actually due to a starspot occultation, and modeled it with a joint starspot-transit model. In the second, we treated it as correlated noise, and modeled it with a Gaussian process. 

\subsection{Starspot modeling}\label{whitesp}
If the distortion of the transit profile is due to a starspot then, other than needing to be corrected, it can be used to constrain the stellar and starspot geometry. In particular, the starspot temperature can be measured. If the bump amplitude is wavelength-dependent, moreover, it would make the starspot scenario more likely and a starspot model would allow a better constraint of its brightness across the WFC3 spectral window. 

\begin{table*}[htb]
\begin{center}
\caption{Transit and spot parameters from MCMC optimization on the white light curves. In the second column, the values in parentheses are the maximum-likelihood values, used in the fit of the spectroscopic channels (Section \ref{corrsyst}).}
\label{whiteres}
{\renewcommand{\arraystretch}{1.8}
\begin{tabular}{llll}
\hline                                     
\hline
Parameter                                       & Spot-free transit model & Spot-transit model & GP \\
\hline
Radius ratio \kr                             &   $0.1662^{+0.0013}_{-0.0018}$ & $0.1675^{+0.0008}_{-0.0017}\,(1.661)$ & $0.1674^{+0.0017}_{-0.0018}$ \\
Transit midpoint $t_0$ [BJD$_\mathrm{UTC} - 2457629.13125$]     &     $0.0702\pm0.0020$ & -$^{(a)}$ & $0.0689^{+0.0032}_{-0.0019}$\\
Planet initial mean anomaly $M$ [$^\circ$]              & - & $256.18^{+0.14}_{-0.63}\, (255.7)$  & -\\
Scaled semi-major axis $a/R_\ast$             & $7.36\pm0.11$ & - $^{(b)}$ & $7.37\pm0.11$ \\
Stellar density [$\rho_\odot$]     & - & $1.68^{+0.10}_{-0.09}\,(1.75)$ & -  \\
Orbital inclination [$^\circ$]                & $85.39\pm0.20$ & $85.31^{+0.19}_{-0.18}\,(85.38)$ & $85.37 \pm 0.20$\\
Linear LD coefficient $u_a$            & $0.25^{+0.08}_{-0.07}$ & $0.18^{+0.08}_{-0.06}\,(0.22)$ & $0.27\pm0.09$\\
Quadratic LD coefficient $u_b$       & $0.18\pm0.09$ & $0.15^{+0.10}_{-0.07}\,(0.22)$ & $0.21^{+0.10}_{-0.09}$\\
Spot longitude $\lambda$ [$^\circ$]        & - & $275.0^{+16.7}_{-4.1}\,(288.6)$ & - \\
Spot latitude $\phi$ [$^\circ$]              & - &$75.8^{+0.6}_{-2.8}\,(73.7)$  &  -  \\
Spot size $\alpha$ [$^\circ$]                & - &$4.1^{+5.3}_{-1.2}\,(5.6)$ & - \\
Spot contrast $f_c$                               & - & $0.46^{+0.14}_{-0.09}(0.37)$& - \\
Reduced $\chi^2$                                 & 1.66 & 1.25 & 1.09 \\  
\hline  
\end{tabular}}
\end{center}
\begin{tablenotes}\footnotesize
    \item $^{(a)}$ The \texttt{KSint} model measures the mean anomaly instead of the transit midpoint.
    \item $^{(b)}$ \texttt{KSint} uses $\rho_\ast$ instead of $a/R_\ast$.
\end{tablenotes}
\end{table*}

In the hypothesis that the transit contains a starspot occultation, the parameters describing the instrumental systematics should be uncorrelated with those describing the starspot, as the systematics occur on a much shorter time scale than stellar rotation and usual starspots lifetimes. To reduce the number of free parameters in the fit, we therefore corrected the band-integrated transit for the best-fit values of the systemtics posterior parameter distributions of the spot-free model. 

Stellar models can be used to estimate the contribution of starspots to the apparent transit depth \citep[e.g.][]{pont2008,pont2013, sing2011,mccullough2014}. A geometric approach can alternatively be used to model the effect of starspots on the whole transit profile, but in a limited spectral window. In this respect, we used \texttt{KSint} \citep{montalto2014} to model the planet and starspot contribution to the transit profile. As this code analytically solves the equations describing the starspot-planet occultation, it is much faster that a numeric code and can be implemented in an MCMC routine. Given our ignorance on possible unocculted starspots in the transit, we modeled only one occulted starspot, rensponsible for the transit distortion. The free parameters in this model were \kr~ and initial mean anomaly of the planet $M$ (required by \texttt{KSint} instead of the transit midpoint), and the starspot longitude $\lambda$, latitude $\phi$, angular size $\alpha$ and brightness contrast $f_c = 1 - F_\mathrm{spot}/F_\star$ (where $F_\mathrm{spot}$ and $F_\star$ are the starspot and stellar fluxes, respectively) with respect to the quiet stellar surface.\footnote{\texttt{KSint} uses the stellar density in place of $a/R_\ast$.} We fitted for the limb darkening coefficients as with the spot-free model. The code uses the same coefficients for both the star and the starspot, which is an acceptable approximation given the precision of the data. We fitted again for the normalization constant $C$ in order to balance the starspot-induced flux baseline variations, as in \texttt{KSint} only the quiet stellar surface has unitary flux. We fixed the rotation period of the star and the orbit eccentricity to \cite{hebrard2013_w52}'s values (Table \ref{parameters}).

By measuring the Rossiter-McLaughlin anomaly, \cite{hebrard2013_w52} measured a $24^{+17}_{-19}$ deg misalignment between the stellar spin axis and the planet orbit axis. Under the assumption of observing the same starspot in consecutive transits, \cite{mancini2017} showed that, instead, the system likely has a low or null misalignment ($3.8 \pm 8.4$ deg). Given the large uncertainties on both measurements, the two results are actually not in strong disagreement. As our observations contain a single transit, we were not able to test any of the published results, and decided to adopt \cite{hebrard2013} misalignment. We implemented this scenario by fixing the stellar axis to form an angle of $85.35^\circ - 24^\circ = 61.35^\circ$ with the plane of the sky (recalling that $85.35^\circ$ is the planet orbit inclination measured in the discovery paper). 

\begin{figure}[hbt]
\epsscale{1.2}
\plotone{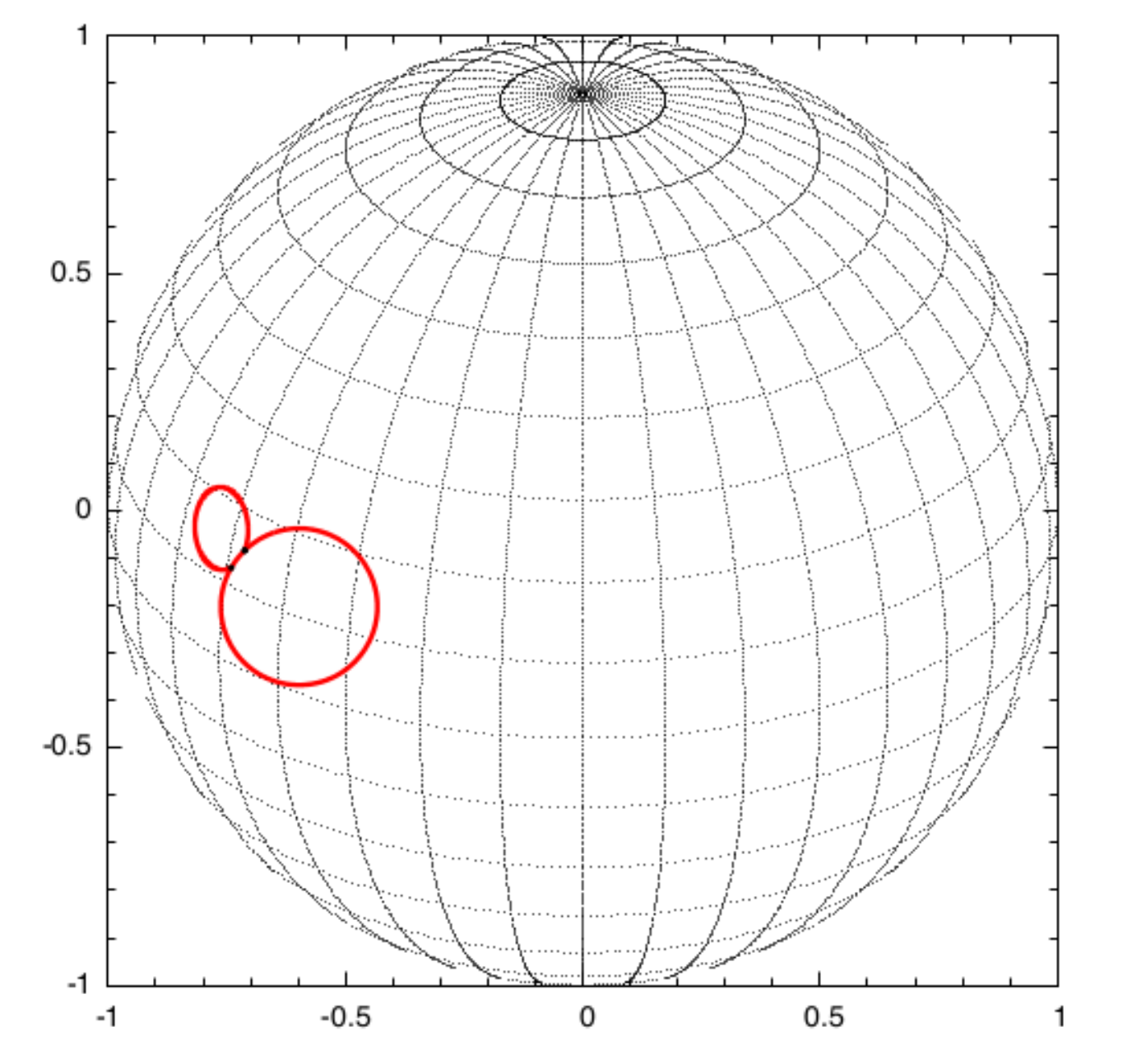}
\caption{Example of geometric configuration of star (unitary radius grid), planet (larger transiting circle), and starspot produced by \texttt{KSint}. The starspot's shape is projected on the stellar disk.}
\label{arcs}
\end{figure}

The starting values for the MCMC optimization of the spot parameters were chosen by examining plots such as Figure \ref{arcs}, which can be produced with \texttt{KSint}. Once the starspot is occulted by the modeled transiting planet, a least-squares minimization followed by an MCMC found the spot's optimal location in order to produce the best-fitting transit geometry. Tens of degrees-wide Gaussian priors were used for the spot coordinates, in order not to affect the derived parameters (Table \ref{priortransits}). We allowed the starspot angular sizes to vary in the range $0^\circ-20^\circ$, in order to consider the typical sizes of sunspots \citep[$\sim 2^\circ$,][]{mandal2017} and of the largest starspots detectable on M dwarfs \citep[$\sim 7^\circ$,][]{berdyugina2011}. 

On the Sun, bright faculae are generally observed in the surroundings of dark spots. If faculae were affecting the transit of WASP-52b, they would be close to the transit center. Faculae tend instead to be observed close to the stellar limb, as also discussed by \cite{mancini2017} for their WASP-52 observations. Moreover, to suppose that a bright spot distorted the transit profile, our spot-free model should have found a significantly larger \kr~ value than other published results, while our result in the near-infrared is within 1-$2\sigma$ of these values. We therefore avoided considering a bright spot in the transit profile and modeled only a dark spot, by imposing a Uniform prior on the brightness contrast going from 0 to 1.
\begin{figure}[htb]
\epsscale{1.25}
\plotone{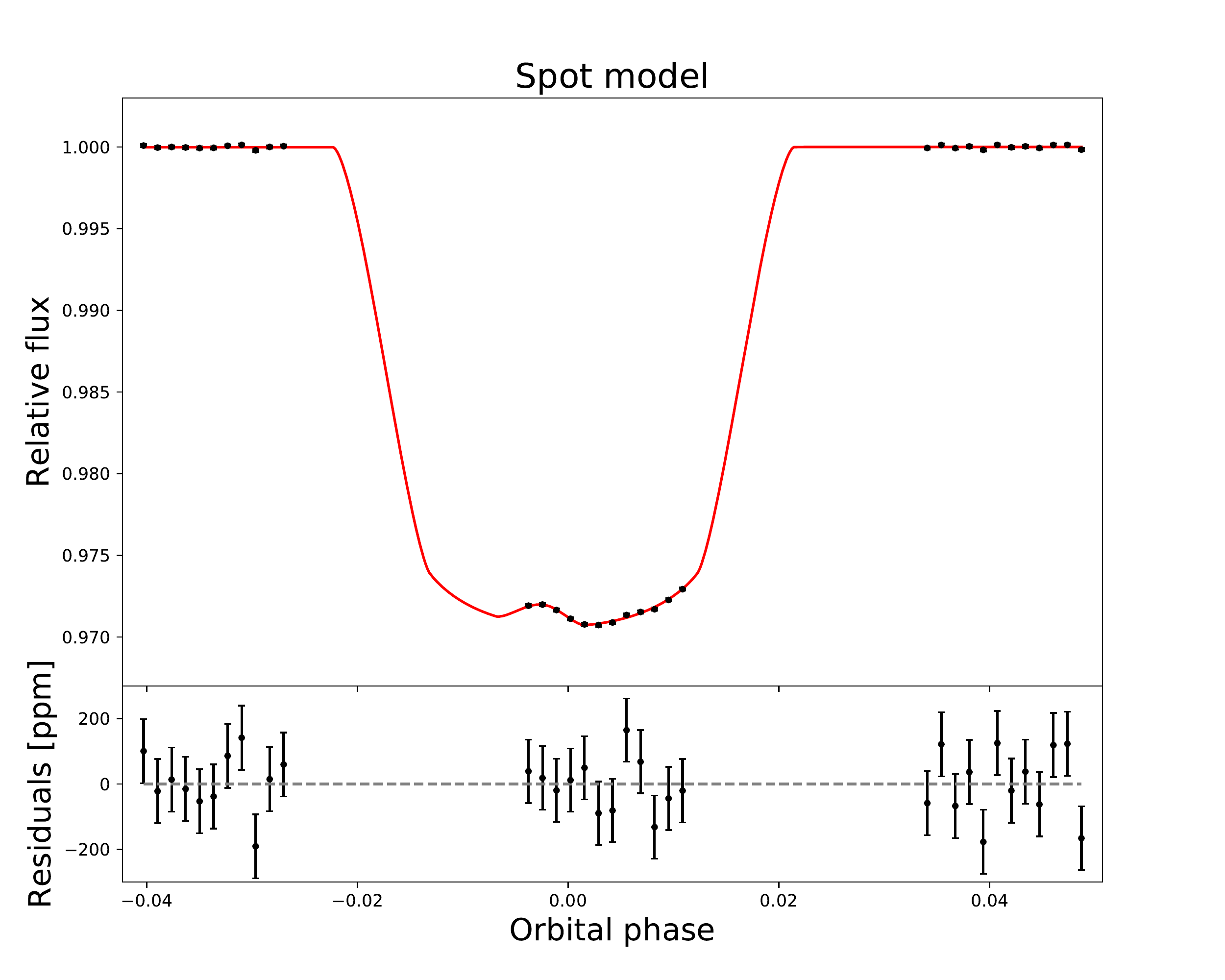}
\caption{Band-integrated transit and best spot-transit model, corrected for the systematics.}
\label{whitetr_spot}
\end{figure}

\begin{figure*}[htb]
\centering
\epsscale{1.2}
\plotone{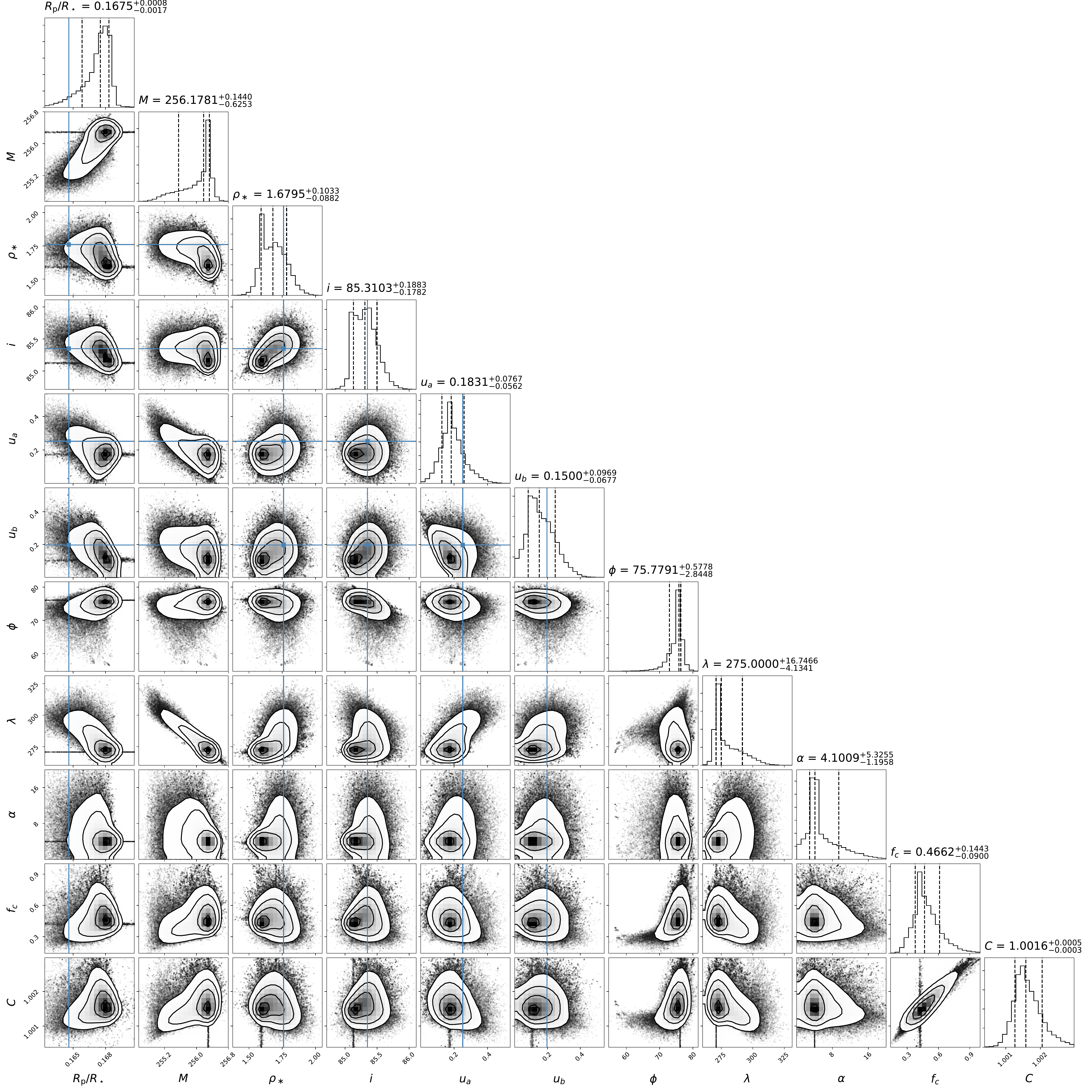}
\caption{Marginalized posterior distributions and correlations for the MCMC with the spot model, showing the 1-, 2- and $3\sigma$ credible intervals. The blue lines and dots show the values reported by \cite{hebrard2013_w52}, as well as the LD coefficients interpolated from stellar models. This plot was realized with the \texttt{corner.py} package \citep{corner}.}
\label{cornerspot}
\end{figure*}

Figure \ref{whitetr_spot} presents the best-fit model and Figure \ref{cornerspot} reports the marginalized posterior distributions and correlations among the parameters of this model. In Table \ref{whiteres}, the mean and $1\sigma$ credible interval values are reported, as well as the best-fit values which were later used for the fit of the spectroscopic channels (Section \ref{corrsyst}). Despite the correlation between starspot size and brightness contrast, considering a simple scenario with a single starspot reduces the risk of incurring in multiple degeneracies which would hamper the physical interpretation of the spot parameters. Instead, we observed a strong correlation between the starspot contrast $f_c$ and the flux normalization $C$. 

By using the black-body law, 
\begin{equation}
    1 - f_c = \frac{ e^{h \nu / k T_{\mathrm{eff}, \star}} - 1}{ e^{h \nu / k T_\mathrm{eff, spot} }-1},
    \label{planck}
\end{equation}
where $h$, $k$, $\nu$, $T_{\mathrm{eff}, \star}$ and $T_{\mathrm{eff}, \star}$ are Planck's constant, Boltzmann's constant, mean WFC3 band frequency, and stellar and starspot effective temperature, respectively. With this expression, we derived a spot effective temperature $T_\mathrm{eff, spot} \simeq 4050^{+370}_{-230}$ K (including the uncertainties on $T_{\mathrm{eff}, \star}$), about $2\sigma$ colder than the temperatures measured by \cite{mancini2017} with their starspot model and compatible with a K star \citep[e.g.][]{berdyugina2005}. If the starspot scenario is correct, such a temperature could hardly allow the presence of water molecules in the starspot. 

\subsection{Transit model with Gaussian process}\label{whitegp}

Here we discuss the possibility that the starspot-like feature can be modeled as correlated noise. This includes the possibility that the transit distortion was actually not produced by stellar ``noise'', but by another type of correlated noise or systematics. Without any further information, we chose a Gaussian process (GP) to represent it in a non-parametric form. We also wanted to investigate whether GPs could be a viable alternative to model starspot occultations, in the case there is actually one in the data set. Despite the option of spot modeling -- the most direct one to attempt reconstructing the starspot distribution on the stellar surface --, this practice quickly becomes computationally demanding when starspots are more than a very low number. Moreover, the combination of high-precision observations and intrinsic degeneracies of the starspot inversion problem make spot modeling less convenient, as the approximation of e.g. using the same limb darkening law for the star and the starspots, or the use of the same limb darkening law for faculae might introduce significant biases in the results. Because of this, non-parametric representations of stellar occultations could be a convenient option for future surveys, as for out-of-transit stellar variations \citep[e.g.][]{haywood2014,aigrain2016,foreman-mackey2017}.

GPs, which enable the modeling of correlated noise, are thoroughly discussed by e.g. \cite{rasmussen-williams2006}. A GP allows cleaning the transit signal from instrumental systematics or other correlated noise, without the need of specifying a function for such noise. The only need is the choice of a function (``kernel'') modeling the covariance between data points. As they are localized events, starspot occultations should be modeled with non-stationary kernels \citep[][Gibson, priv. comm.]{louden2017}, while stationary ones (which depend only on the distance between points) have been successfully employed to model instrument systematics in transit observations \citep[e.g.][]{gibson2012,gibson2013,gibson2013gemini,louden2017}. However, e.g. \cite{louden2017} used a stationary kernel to also model possible starspot occultations in a transit.

We implemented the GP with the \texttt{George} package \citep{george,foreman-mackey2015_GP}, and used a \cite{mandelagol2002} transit as mean function (the same of the spot-free scenario). After numerous tests with the default \texttt{George} kernels and their combinations, we found that a (stationary) Mat\'ern 3/2 kernel and a (non-stationary) cosine kernel best performed to model the \textit{HST} systematics and the starspot-like feature, respectively. The use of GP regression to model the systematics, instead of the previously fitted analytic model, had the goal of investigating the performance of this relatively assumptions-free technique for our case.

For two times $t_i$ and $t_j$, the covariance was modeled as
\begin{equation}
\begin{split}
 k(t_i, t_j) & = \bigg( 1 + \sqrt{3} \, \frac{|t_i - t_j|}{\eta}  \bigg) e^{-\sqrt{3} \, |t_i - t_j|/\eta}  \\
             & \cdot \cos \bigg( \frac{2 \pi |t_i - t_j|}{P} \bigg) + \delta_{ij} \sigma_w^2,
\end{split}
\label{eqmatern}
\end{equation}
where the parameters $\eta > 0$, $P > 0$ and $\sigma^2_w$ represent the inverse scale length, the period of the GP and the average variance of the data points, respectively, and $\delta$ is the Kronecker delta. The parameter $\sigma_w$ allowed to scale the uncertainties by adding a diagonal term in the covariance matrix, but fitting for this parameter resulted in overfitting, i.e. $\tilde {\chi}^2 < 1$. Because of this, and because we did not observe any variation of the uncertainties with time, we fixed $\sigma^2_w$ in our analysis.

\begin{figure}[htp]
\epsscale{1.25}
\plotone{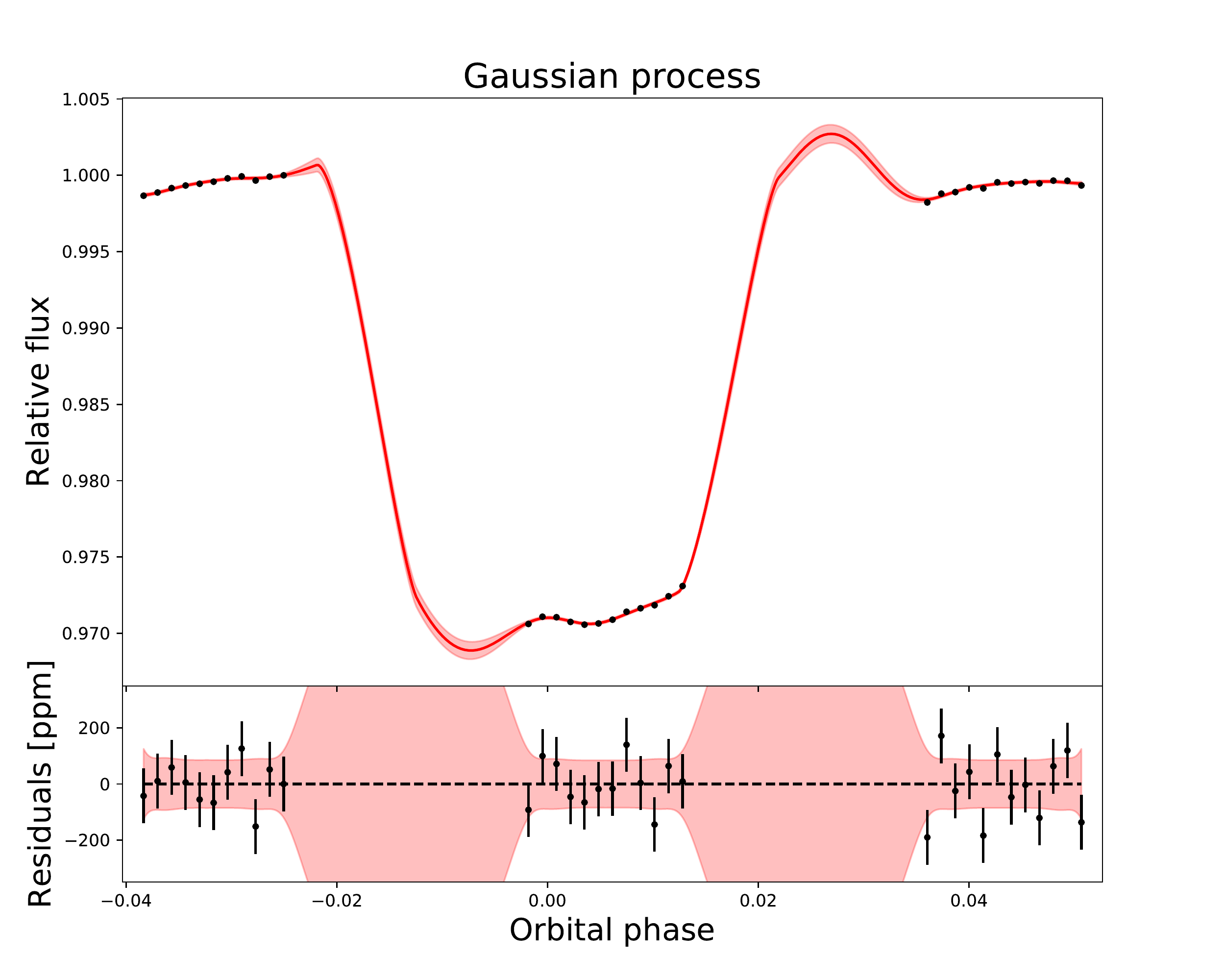}
\caption{Band-integrated transit and best GP model. The predicted $1\sigma$ uncertainty of the GP is shown in light red. Note the larger uncertainty on the edges of the transit, where there are no data points.}
\label{gp_white}
\end{figure}

We used the likelihood function automatically calculated by \texttt{George} for the least-squares minimization and adopted the gradient of the likelihood function in the optimization algorithm. In the MCMC analysis, we set Gamma distributions with shape parameter unity for the priors on $\eta$ and $P$, following \cite{gibson2012}. These are described by
\begin{equation}
p(x) = \left\{
\begin{array}{ll}
0 & ~\mathrm{if}~x < 0 \\
\dfrac{1}{l} e^{-x/l} & ~\mathrm{if}~x \geq 0 \\
\end{array}
\right.,
\end{equation}
where $x$ is the hyperparameter and $l$ the length scale of the hyperprior. A length scale of $200$ and $10^{-3}$ for $\eta$ and $P$, respectively, were appropriate for our data set. \cite{gibson2012} used these same length scales for a Mat\'ern 3/2 and constant kernel hyperparameters, respectively. We used the same Gaussian priors for the transit parameters used in the previous cases. All priors are presented in Table \ref{priortransits}.

\begin{figure*}[hbt]
\centering
\epsscale{1.1}
\plotone{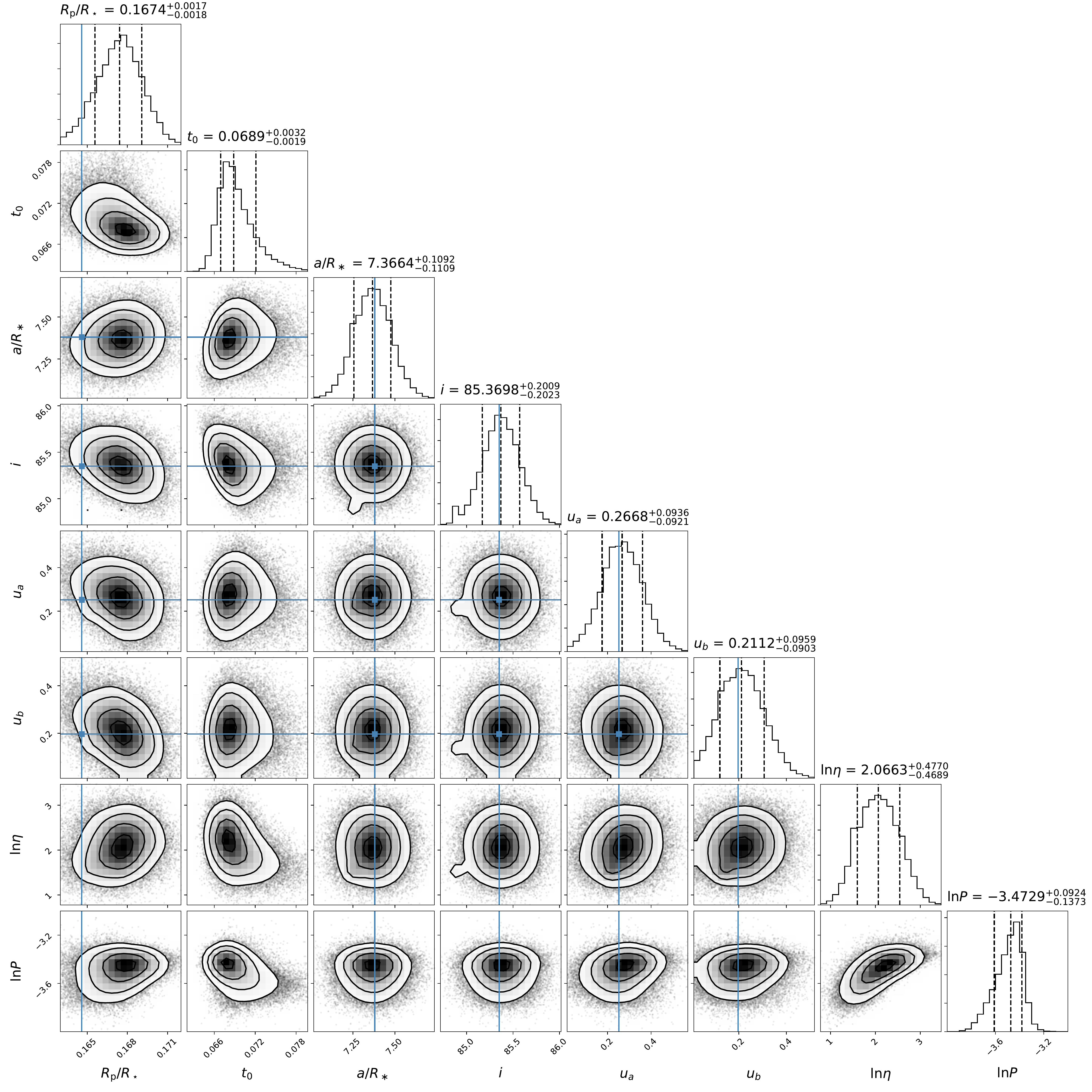}
\caption{Marginalized posterior distributions, 1-, 2- and $3\sigma$ credible intervals, and correlations for the MCMC with the GP model. The blue lines and dots show the values reported by \cite{hebrard2013_w52} and LD coefficients interpolated from stellar models.}
\label{corner_gp}
\end{figure*}

We present the best-fit GP model in Figure \ref{gp_white}, while the marginalized posterior distributions, correlation plots and derived parameters are shown in Figure \ref{corner_gp} and Table \ref{whiteres}. Specifically, the transit midpoint $t_0$ found by the GP occurs $\simeq 2$ min earlier than for the spot-free model. The two values are well within $1\sigma$ uncertainties of $\sim 3$ min.

\section{Spectroscopic transits}\label{spectrotr}

\subsection{Transit modeling}\label{corrsyst}
The preparation of the spectroscopic transits was based on the assumption that the systematics and the noise observed on the band-integrated transit are wavelength-independent and therefore present the same properties in each channel. Each approach used for the fit of the white light curve offered a different way of correcting the spectroscopic transits for these effects.
\begin{enumerate}
    \item The residuals of the spot-free models were used to correct the spectroscopic channels via a ``common-mode'' approach \citep{stevenson2014_w12}, where each transit was divided by the residuals of the best-likelihood model of the band-integrated transit model. Here, the possible starspot occultation, previously masked, was corrected in the spectroscopic channels as a residual from the previous fit.
    \item The spectroscopic channels, corrected for the systematics as in the white light spot model, were not further corrected. All the starspot parameters were fixed to their maximum-likelihood values (Table \ref{whiteres}) but the contrast ratio, which was fitted in an attempt of better constraining the temperature of the possible starspot.
    \item For the approach with the GP, the covariance function of the best-fit model was used to compute an analytic formulation of the combined effect of systematics, stellar noise and remaining red noise. Each spectroscopic channel was then corrected by using this expression. 
\end{enumerate}

\begin{figure}[htb]
\epsscale{1.27}
\plotone{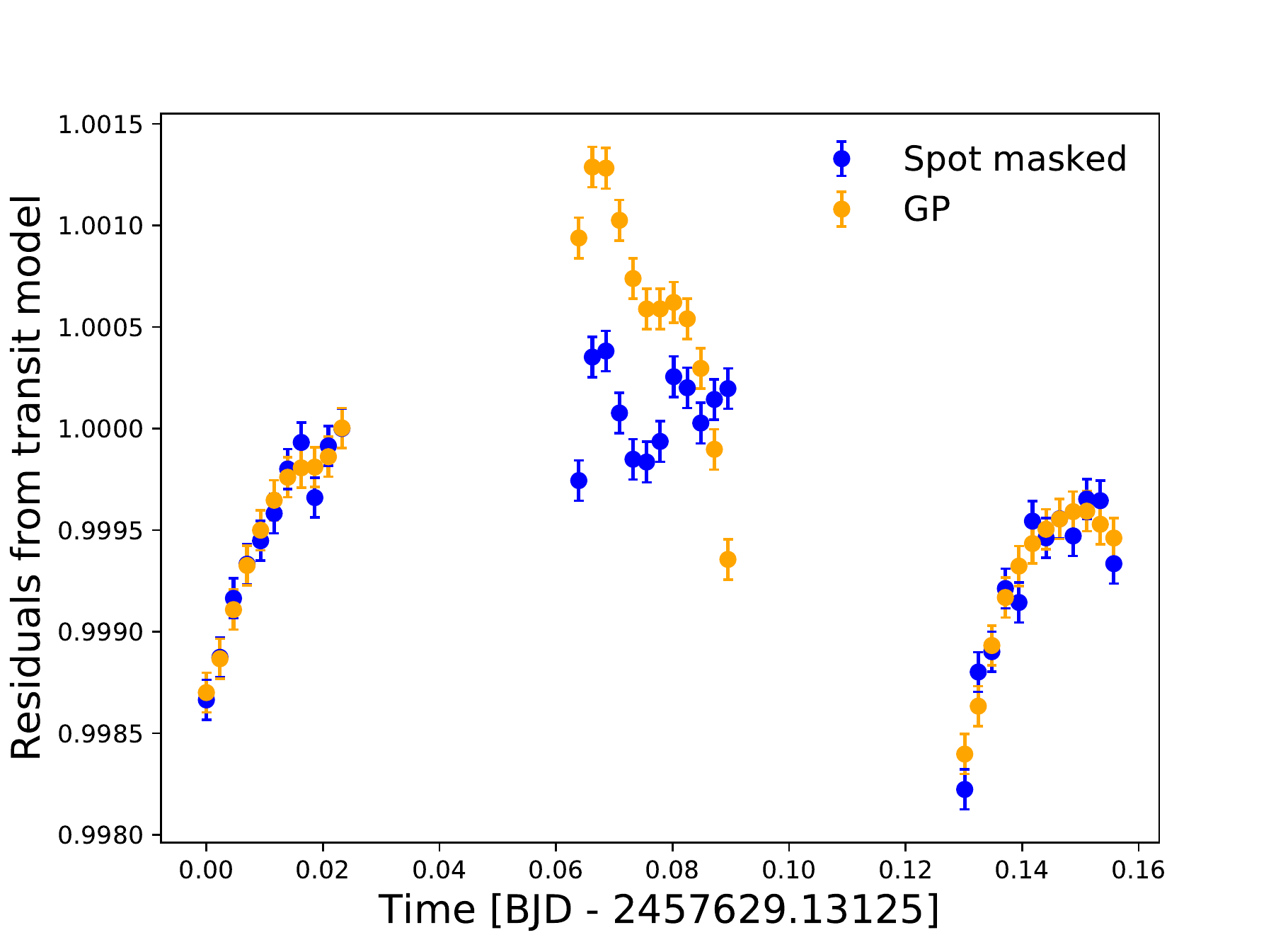}
\caption{Comparison of the systematics and starspot corrections derived from the spot-free and the GP model. The different correction factor at the center of the transit is due to the different band-integrated \kr~ measurement.}
\label{comp_residuals}
\end{figure}

Figure \ref{comp_residuals} compares the ``common-mode'' residuals and the correction factor derived by the GP covariance function. We remark an average $\sim 10^{-3}$ difference in the center of the transit, which is related to different ways of incorporating the normalization factor (the ``$C$'' parameter) in the models, and which is reflected in the differences in \kr~ in Table \ref{whiteres}. This might be better constrained, especially for the GP model, with observations of the out-of-transit light curve, which in a physical sense contain information on unocculted starspots. Such long-baseline observations would be a little convenient use of the telescope time, especially for space-borne instruments such as \textit{JWST}, but simultaneous ground-based follow-ups might provide with some useful constraints. At present, however, we have no information on which model -- whether the spot-masked or spot model, or the GP -- provides a more realistic measure of the white light curve \kr.

For the MCMC on each spectroscpic channel resulting from the spot-free and GP model, we used a standard transit model where \kr~ was set as a free parameter (starting from the best-fit values on the white light curve) and the transit midpoint $t_0$ was fixed to the band-integrated light curve best fit. For the spot model, we correspondingly fixed the mean anomaly $M$. As we expected the starspot constrast to vary with wavelength in a similar way to a black-body function \citep[e.g.][]{berdyugina2005,ballerini2012}, we only left this parameter free, and fixed the starspot coordinates and sizes to their best-fit values. Specifically, we approximated the starspot as a black body, and used Equation \ref{planck} to predict the value of $f_c$ as a function of wavelenght, based on the value obtained from the white light curve best fit. The $f_c$ fitted on the channels was then controlled with a Gaussian prior having the predicted value as mean, and 0.1 as standard deviation. A linear slope was used to model visit-long trends left uncorrected after the white light curve fit. We interpolated the LD coefficients with stellar models as previously described, obtaining the values reported in Table \ref{limbdtable}. Because of the lower signal-to-noise ratio of the spectroscopic transits with respect to the broadband one, the coefficients were fixed during the MCMC analysis.

\begin{table}[htb]
\begin{center}
\caption{Limb darkening coefficients for WASP-52, fixed in the fit of the spectroscopic channels. The first column of each table indicates the wavelength range corresponding to the channel. The second and third columns are for the linear ($u_a$) and quadratic ($u_b$) coefficients.}
\label{limbdtable}
\begin{tabular}{c|c|c}
\hline
\hline
$\lambda$ [$\mu$m] &  $u_a$ & $u_b$ \\
\hline
1.115--1.143  &  0.312   &  0.152 \\
1.143--1.171  &   0.314   &  0.153  \\
1.171--1.199  &   0.297   &  0.150  \\
1.199--1.227   &  0.304   &  0.160  \\
1.227--1.256   &  0.299   &  0.165  \\
1.256--1.282  &   0.286  &   0.174  \\
1.282--1.310   &  0.275  &   0.184  \\
1.310--1.338  &   0.275  &   0.183  \\
1.338--1.366  &   0.266   &  0.192  \\
1.366--1.394   &  0.255  &   0.202  \\
1.394--1.422   &  0.244  &   0.211  \\
1.422--1.450   &  0.239  &   0.218  \\
1.440--1.478   &  0.229  &   0.215  \\
1.478--1.506    & 0.207  &   0.234  \\
1.506--1.534  &   0.208  &   0.234  \\
1.534--1.561   &  0.183  &   0.247  \\
1.561--1.589   &  0.169  &   0.243  \\
1.589--1.617   &  0.158  &   0.246  \\
1.617--1.645   &  0.150  &   0.247  \\
\hline
\end{tabular}\\
\end{center}
\end{table}

\begin{figure}[htb]
\centering
\epsscale{1.2}
\plotone{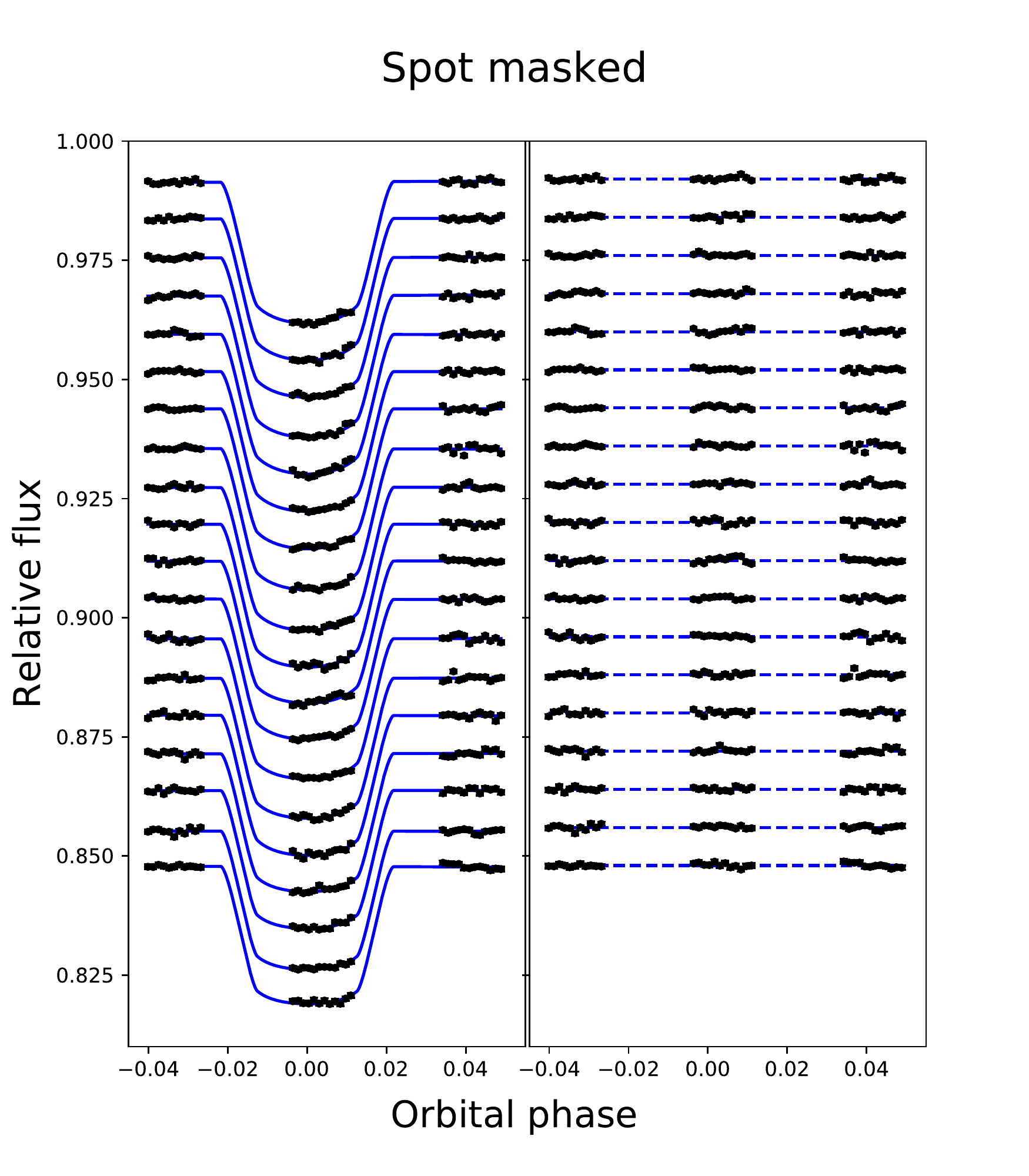}
\caption{Spectroscopic light curves after masking the spot and correcting for the systematics, with relative best fits, and shifted for clarity. The right panel shows the residuals.}
\label{channels_masked}
\end{figure}

\begin{figure}[htb]
\centering
\epsscale{1.2}
\plotone{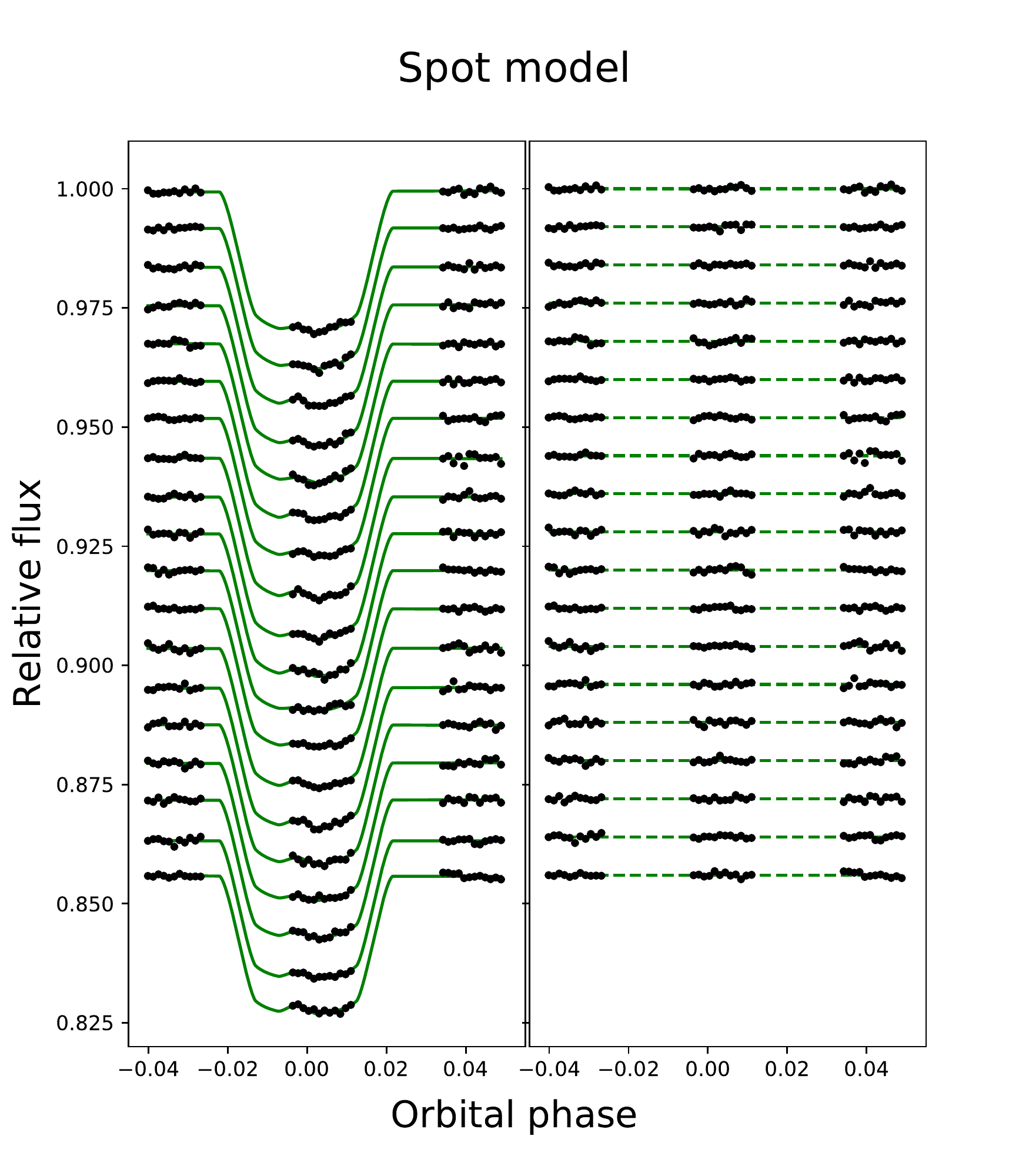}
\caption{Same as Figure \ref{channels_masked}, for the spot-transit model.}
\label{channels_spotmodel}
\end{figure}

\begin{figure}[htb]
\centering
\epsscale{1.2}
\plotone{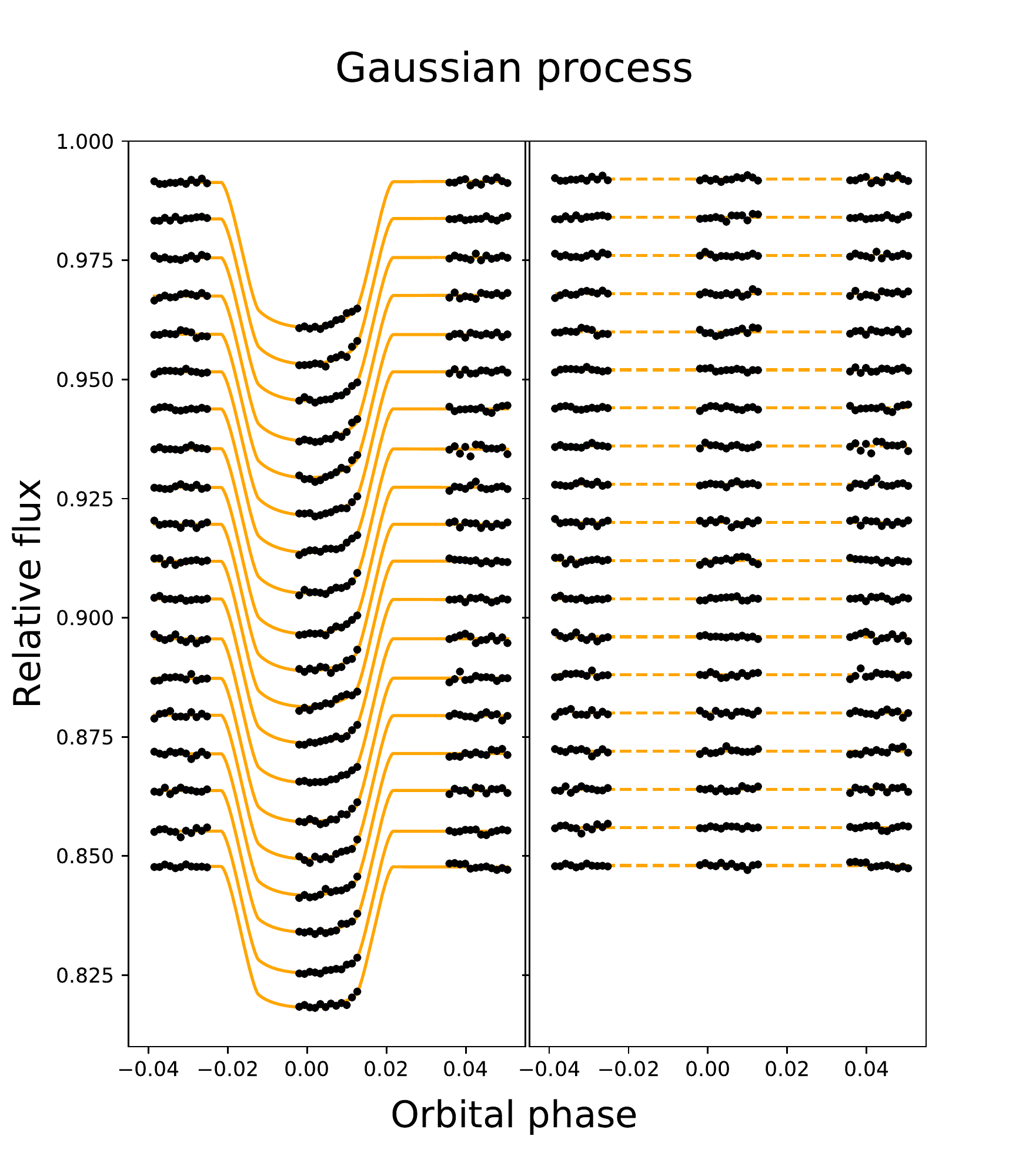}
\caption{Same as Figure \ref{channels_masked} and \ref{channels_spotmodel}, for the GP model.}
\label{channels_gp}
\end{figure}

The fitted \kr~ are reported in Table \ref{kr_compared}, and Figures \ref{channels_masked}, \ref{channels_spotmodel} and \ref{channels_gp} present the best-fit models of the three analyses. In the residuals of Figures \ref{channels_masked} and \ref{channels_gp}, we observe that the starspot-like feature does not vary significantly among the channels, so that we can safely assume the transit distortion to be wavelength-independent at the data level of precision.

\begin{table*}[tb]
\begin{center}
\caption{Compared \kr~ of the spectroscopic channels from the different models. From left to right, spectral window, standard transit model with starspot masked, transit-starspot model and GP regression model. The $D$ parameter expresses the significance of the water absorption feature according to \cite{stevenson2016}'s classification, and the standard deviation of the normalized residuals (SDNR) is taken from the cumulative distribution on all channels.}
\label{kr_compared}
\begin{tabular}{ c | c | c | c }
\hline
\hline
$\lambda$ [$\mu$m]    & Spot-free transit model & Spot-transit model & GP \\
\hline
1.115--1.143  & $ 0.16682 \pm 0.00047 $  &  $ 0.16641 \pm 0.00049 $ & $ 0.16950 \pm 0.00047 $  \\
1.143--1.171  & $ 0.16690 \pm 0.00044 $	 &  $ 0.16647 \pm 0.00052 $   & $ 0.16947 \pm 0.00046 $ \\
1.171--1.199  & $ 0.16597 \pm 0.00042 $  &  $ 0.16594 \pm 0.00046 $  & $ 0.16841 \pm 0.00044 $ \\
1.199--1.227  & $ 0.16695 \pm 0.00041 $  &  $ 0.16666 \pm 0.00045 $ & $ 0.16945 \pm 0.00042 $ \\
1.227--1.256  & $ 0.16588 \pm 0.00042 $  &  $ 0.16542 \pm 0.00045 $ & $ 0.16843 \pm 0.00044 $ \\
1.256--1.282  & $ 0.16592 \pm 0.00038 $  &  $ 0.16593 \pm 0.00044 $  & $ 0.16837 \pm 0.00040 $  \\
1.282--1.310  & $ 0.16609 \pm 0.00042 $  &  $ 0.16585 \pm 0.00046 $ & $ 0.16851 \pm 0.00043 $ \\
1.310--1.338  & $ 0.16674 \pm 0.00041 $  &  $ 0.16674 \pm 0.00044 $ & $ 0.16921 \pm 0.00041 $ \\
1.338--1.366  & $ 0.16791 \pm 0.00040 $	 &  $ 0.16763 \pm 0.00046 $ & $ 0.17038 \pm 0.00042 $ \\
1.366--1.394  & $ 0.16790 \pm 0.00042 $	 &  $ 0.16788 \pm 0.00046 $ & $ 0.17035 \pm 0.00041 $ \\
1.394--1.422  & $ 0.16741 \pm 0.00043 $  &  $ 0.16680 \pm 0.00045 $  & $ 0.16986 \pm 0.00043 $ \\
1.422--1.450  & $ 0.16635 \pm 0.00040 $	 &  $ 0.16600 \pm 0.00042 $   & $ 0.16880 \pm 0.00041 $  \\
1.440--1.478  & $ 0.16652 \pm 0.00042 $  &  $ 0.16654 \pm 0.00048 $  & $ 0.16891 \pm 0.00044 $ \\
1.478--1.506  & $ 0.16647 \pm 0.00045 $  &  $ 0.16654 \pm 0.00051 $  & $ 0.16887 \pm 0.00046 $ \\
1.506--1.534  & $ 0.16650 \pm 0.00044 $	 &  $ 0.16636 \pm 0.00050 $   & $ 0.16888 \pm 0.00045 $ \\
1.534--1.561  & $ 0.16545 \pm 0.00046 $  &  $ 0.16509 \pm 0.00049 $  & $ 0.16783 \pm 0.00046 $ \\
1.561--1.589  & $ 0.16569 \pm 0.00045 $  &  $ 0.16568 \pm 0.00051 $  & $ 0.16805 \pm 0.00047 $ \\
1.589--1.617  & $ 0.16597 \pm 0.00044 $  &  $ 0.16592 \pm 0.00052 $   & $ 0.16820 \pm 0.00046 $  \\
1.617--1.645  & $ 0.16536 \pm 0.00050 $	 &  $ 0.16562 \pm 0.00054 $ & $ 0.16766 \pm 0.00050 $  \\
\hline
  $D$   & $0.93 \pm 0.26$		   &	$0.86 \pm 0.29$	  &   $0.93 \pm 0.27 $ \\
  SDNR [ppm]  & 412  &   412 & 417 \\
\hline
\end{tabular}
\end{center}
\end{table*}

\begin{figure}[htp]
\epsscale{1.25}
\plotone{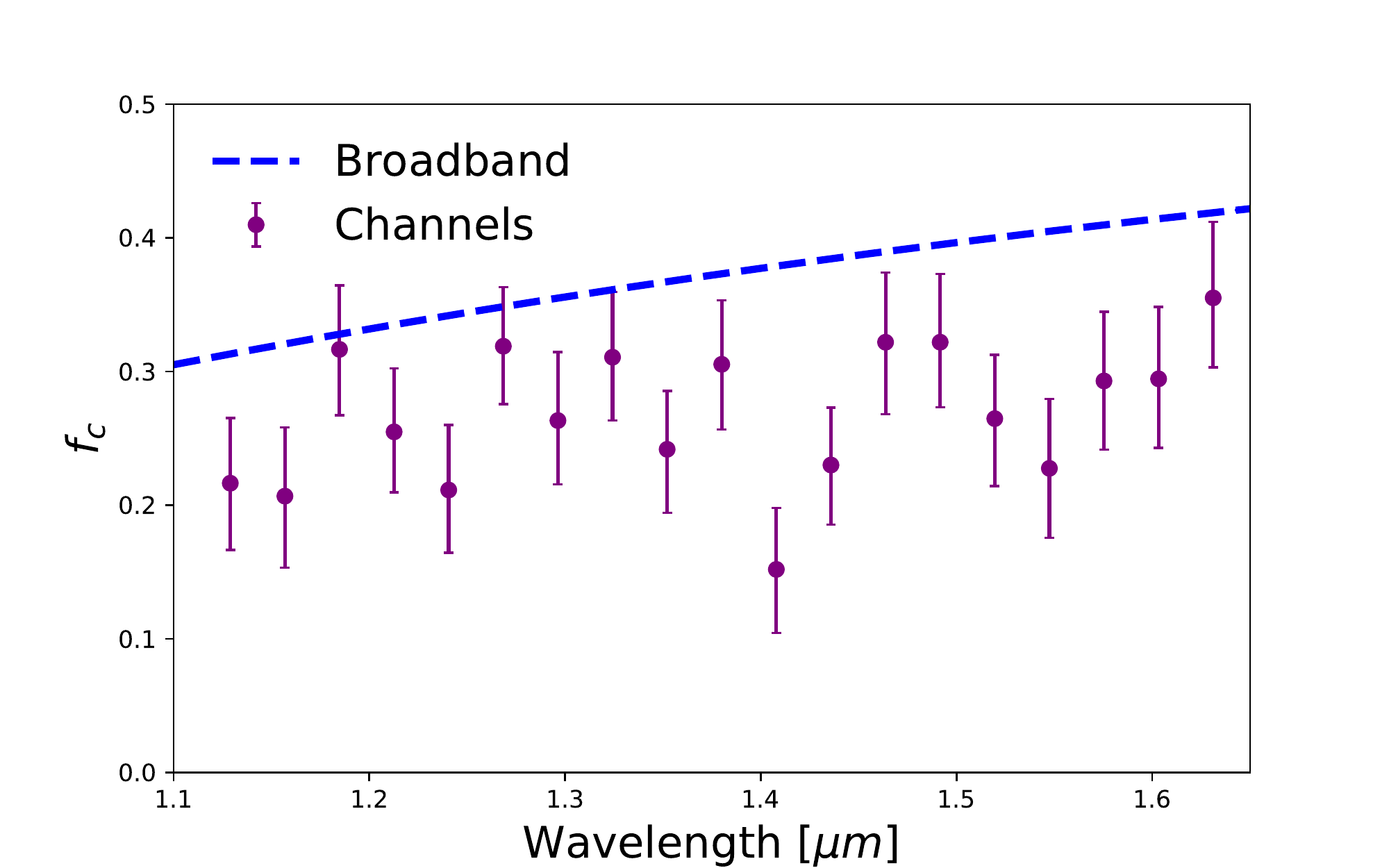}
\caption{Mean and $1\sigma$ credible interval of the starspot contrast ratio predicted from the maximum-likelihood fit on the band-integrated light curve (dashed blue line), and measured on the spectroscopic transits (purple points with error bars).}
\label{contrast_pred}
\end{figure}

This has implications for the fit of the contrast ratio of the possible starspot. Figure \ref{contrast_pred} presents the fitted brightness contrasts and the prediction of the starspot contrast across the WFC3 channels, based on the maximum-likelihood white light curve fit as previously described. At the level of precision of the WFC3 data, no dependence of the contrast with wavelength was recovered. This might be due to a too faint starspot signal across the spectroscopic channels, or to the fact that the starspot-like feature was not actually produced by a starspot. Moreover, the $f_c$ recovered on the spectroscopic channels is consistently lower than the prediction from the white light curve. This might be explained by the fact that our prediction was based upon the assumption that a starspot can be approximated as a black body, while our problem requires the more accurate use of stellar models. If the occultation scenario is correct, but the contrast difference is below the noise level, our joint ignorance of the starspot temperature and of the fractional coverage in starspots of the stellar surface prevents us from estimating the amount of contamination of the water feature \citep{fraine2014}. 

\begin{figure}[htp]
\epsscale{1.2}
\plotone{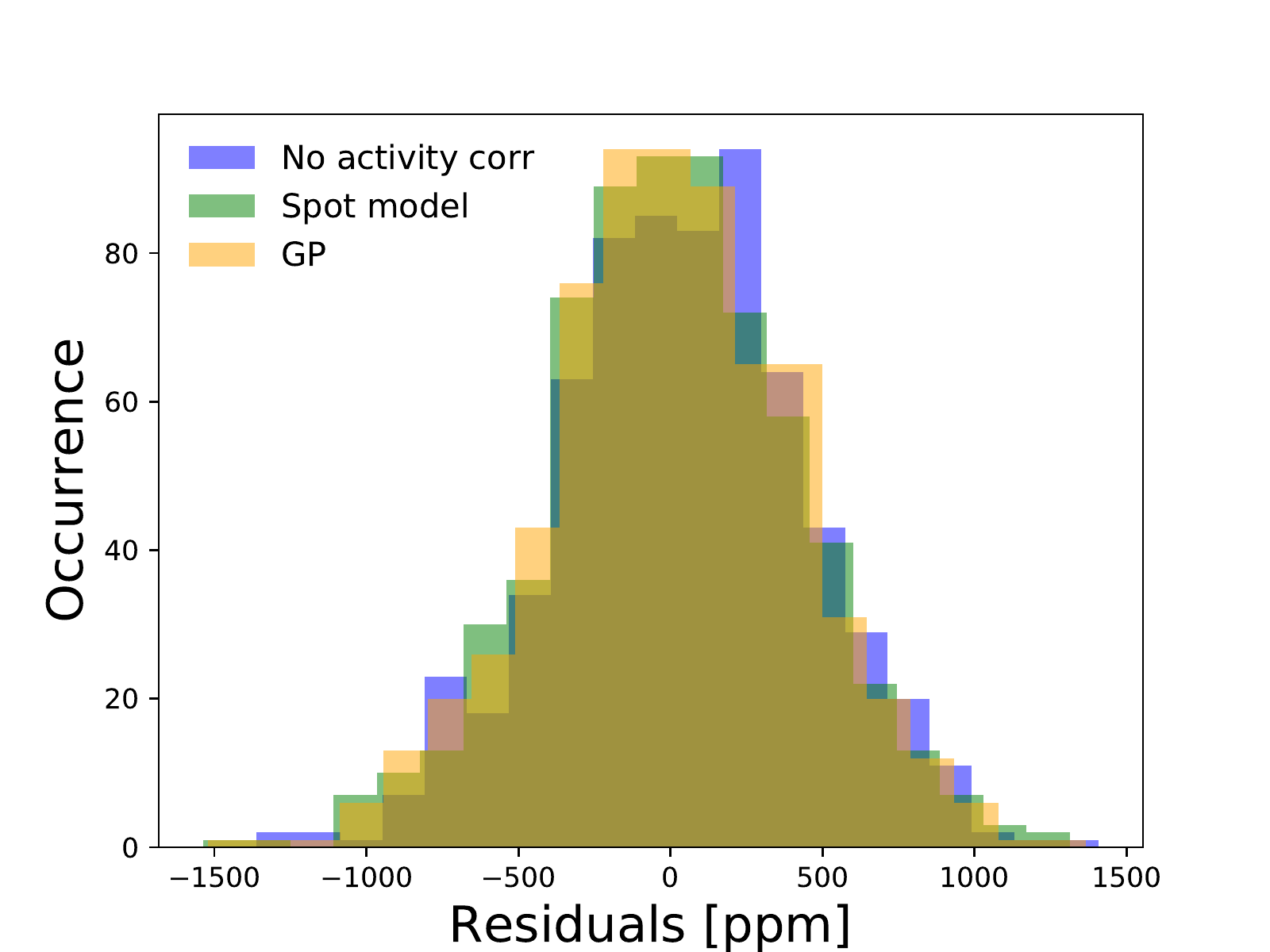}
\caption{Cumulative histogram of the residuals on the spectroscopic channel fits from the three analyses.}
\label{sdnr}
\end{figure}

In Figure \ref{sdnr}, we present the cumulative histograms of the residuals on the spectroscopic channels from the three analyses, in order to inspect for the presence of systematic biases in the three methods. No significant difference appears from this plot, which also shows the Gaussian distribution of the residuals -- an indication that correlated noise features, such as the starspot-like feature, do not significantly affect our measurements.

\subsection{Transmission spectrum}

\begin{figure*}[htb]
\epsscale{1.0}
\plotone{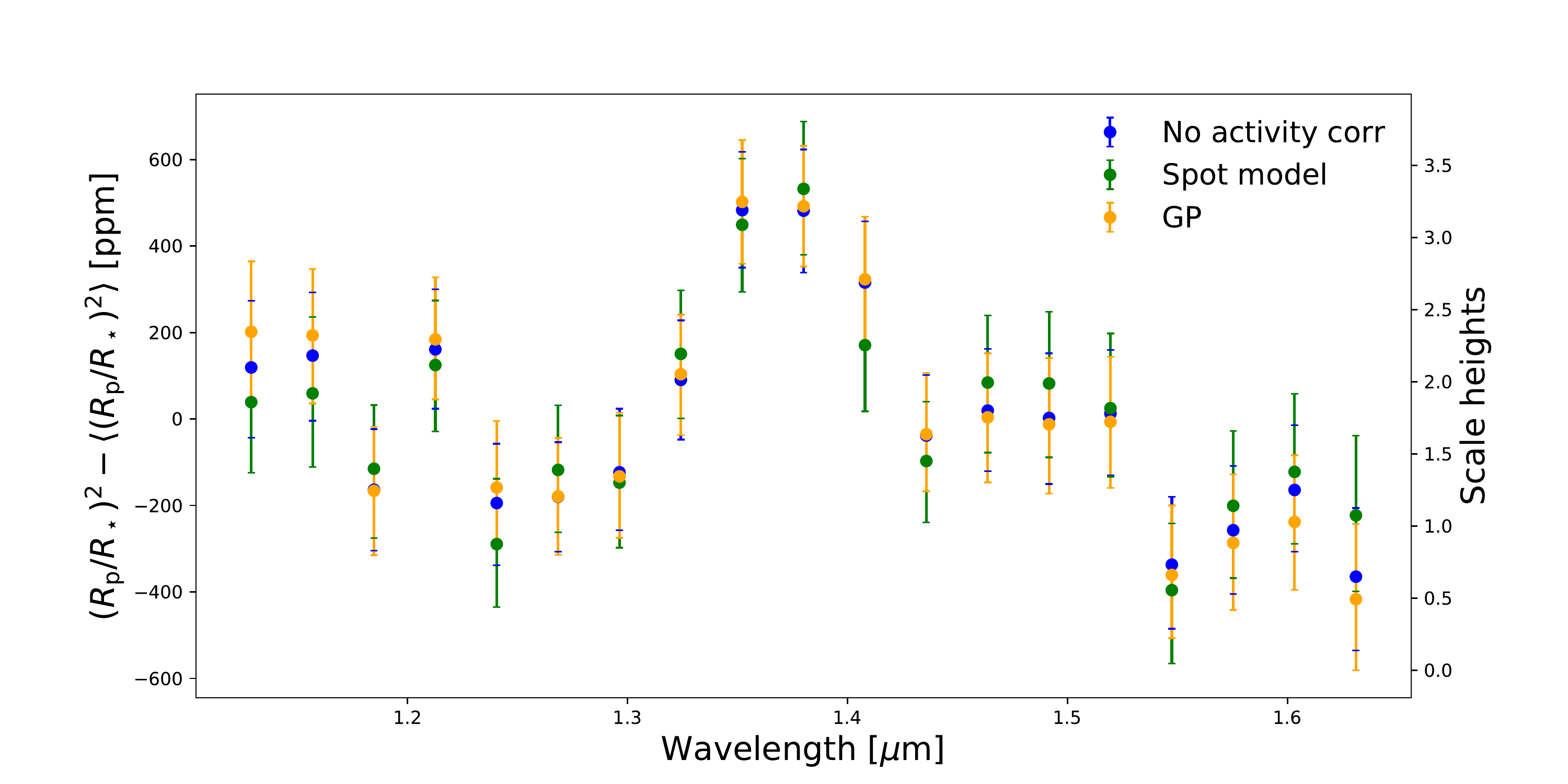}
\caption{Transmission spectra of WASP-52b from the spot-free, spot-transit, and GP model. The mean flux value was removed from each spectrum to allow an easier comparison.}
\label{spectra_comp}
\end{figure*}

Figure \ref{spectra_comp} compares the transmission spectra obtained from the different analyses. The mean value of each spectrum -- which changes among the different analyses because of the different white light curve \kr -- was removed in order to ease the comparison. Different reference \kr~ were expected from the considerations in Section \ref{corrsyst} and produced no difference on the shape of the transmission spectra, all of which provide a robust detection of water absorption at $1.4 \, \mu$m.

The water absorption significance was estimated with the \htoj\ index, where the measure was referred to the average transit depth between 1.36 and 1.44 $\mu$m \citep{stevenson2016}. For the J band, i.e. the baseline, we used the 1.24-1.30 $\mu$m wavelength range. In Table \ref{kr_compared} we present the indices resulting for each approach, which gave a weighted average of $D = 0.91 \pm 0.30$. According to \citeauthor{stevenson2016}'s classification, this value indicates the likely presence of obscuring clouds in the atmosphere of the planet, in agreement with published results \citep{kirk2016, louden2017}.

\subsection{Model comparison}

All the parameters derived with the three methods on the band-integrated transits (Table \ref{whiteres}) are at $\sim 1\sigma$ agreement. The performance of each analysis was evaluated using the reduced chi-square ($\tilde{\chi}^2$, Table \ref{whiteres}) on the band-integrated transits and the standard deviation of the normalized residuals (SDNR, Table \ref{kr_compared}) on the spectroscopic channels \citep[e.g.][]{campo2011, stevenson2012}. The Bayesian or the Akaike information criteria \citep{schwarz1978, akaike1974} are not suitable indicators for our case, as they do not allow comparing models which use subsets of the same data set (i.e., different numbers of data points), excluding therefore the case where the spot feature was masked.

Using $\tilde{\chi}^2$, specifically, allowed control on the possibility of overfitting by the GP. Even if GPs intrinsically weight the goodness-of-fit by penalizing the complexity of the fitted model \citep[e.g.][]{rasmussen-williams2006,gibson2014}, we found that some combinations of kernels and fitted hyperparameters tended to achieve $\tilde{\chi}^2 < 1$. This estimator showed that, by fixing the white noise scaling parameter $\sigma_w$, the GP fitted the data set with a slightly better performance than the spot model, without overfitting. Even if the GP model is favored over the other models, both in terms of  $\tilde{\chi}^2$ and of lower number of free parameters, the spot model yields a more complete \textit{physical} description of the possible configuration of the system, which neither the masking nor the GP can yield.

The SDNR informs on the efficacy of the different strategies in correcting the spectroscopic channels for both instrument systematics and starspot-like feature. The difference in the respective indicators is at most 5 ppm among the three analysis methods. This means the analyses can be considered equally effective, as is confirmed from the resulting transmission spectra.

\section{Discussion}\label{discussion}
Starspot crossings are not generally considered an issue in infrared transmission spectroscopy, because of the negligible contribution of their black-body emission compared to the stellar one. In this work, we presented the possible detection of a starspot occultation in near-infrared observations of WASP-52 and discussed different ways of correcting the spectroscopic transits before deriving the transmission spectrum. We found that the transit distortion is only likely to affect the reference radius of the planet, much like an unocculted starspot in the visible.

With high-resolution, continuously observing instruments such as \textit{JWST}, masking starspot occultations might become a less convenient strategy than with \textit{HST}. The focus on cool, active stars identified by upcoming surveys such as \textit{TESS} will result in more starspot occultations, whose masking will hide relevant information on the transit parameters. While only starspot models allow the reconstruction of the occultation geometry, the computational cost of comparing multiple scenarios across long baselines could become prohibitive. In this context, the use of non-parametric models such as Gaussian processes (GPs) can be explored as a viable alternative.

In search of an optimal strategy for correcting the spectrum from the possible starspot occultation, we compared the implications of masking the transit distortion and of modeling it both with a starspot model and with a GP. We showed that the transit distortion is wavelength-independent at the precision of our data. With a geometric model approach, we obtained a $\simeq 4050^{+370}_{-230} \, \mathrm{K}$  fit on the spot temperature, which makes the possibility of stellar water contamination in the transmission spectrum unlikely. As WASP-52A is an active star, the possibility that other, colder non-occulted starspots could be present on the stellar disk also cannot be excluded. As we found no dependence of the spot contrast among spectroscopic channels, however, the possibility that the transit distortion is not due to a starspot remains valid. 

Despite this, given the agreement of all reduction methods, the water absorption feature in the \textit{HST}/WFC3 G141 spectrum of WASP-52b is unlikely to be affected by the starspot-like distortion. The \cite{stevenson2016} indicators of the significance of the water feature on the three resulting spectra are all in agreement, supporting a robust water feature detection which is likely muted by aerosols in the atmosphere of the planet. While our analysis showed that different approaches can result in $\lesssim 1\sigma$ variations of the spectrum baseline, the uncertainty in the reference planet radius, and therefore in the spectrum baseline, can likely be taken into account in retrieving the atmospheric state by the use of a scaling factor \citep[e.g.][]{benneke2012, line2013}. 

We found similar uncertainties on the transit depth from the three analyses. Moreover, the GP model resulted in a transit midtime $\lesssim 2$ minutes earlier than the model in which the spot is masked (but still within the uncertainties of $\sim 3$ minutes). To our knowledge, this is the first time a non-stationary GP kernel is used for the specific purpose of modeling a possible starspot occultation in a transit profile. This promising result needs to be further investigated, and can find applications both in transmission spectroscopy and broadband photometry. In particular, stationary kernels could be the most convenient choice for modeling other forms of stellar noise which occur on shorter time scales than the transit, such as granulation, which happens over tens of minutes and can affect the apparent \kr~ \citep{chiavassa2017}.

Higher-precision instruments such as those aboard \textit{JWST}, thanks also to their ability of completely covering the transit profile, will likely provide additional constraints in these kinds of scenario. As spot models over large data sets, GPs become computationally demanding when applied on $\gtrsim1000$ points, because of the need of iterating large matrix inversions. \cite{gibson2013gemini} proposed to use the maximum likelihood type II method in the MCMC exploration, i.e. to fix the hyperparameters to their least-squares best value, in order to reduce the computation time. With our small data set of a few tens of data points per transit, this was not an issue, while with \textit{JWST} observations it might affect the choice of the most convenient approach.  

WASP-52A is a $\sim 5000$ K star, but starspots in colder stars might contribute more significantly to the water feature in a transmission spectrum than what was found with this study. Predicting how activity features will affect \textit{JWST} observations -- and the computation cost of using GPs in place of starspot modeling -- requires an analysis of multiple transit shapes and of varying sizes and temperatures for the occulted starspots, which is beyond the scope of this paper. We plan to carry out a study of this nature with the use of synthetic data. We expect a stronger relative impact in \textit{JWST} than in \textit{HST} observations, as error bars on the transmission spectrum will be smaller than the few 100 ppm observed for this 10.1 mag$_H$ star.

Many future studies will focus on WASP-52b. Combined \textit{HST}/STIS observations obtained from the Panchromatic Comparative Exoplanetology Treasury (PanCET) program (GO 14767, PIs Sing \& L\'{o}pez-Morales) and \textit{Spitzer}/IRAC observations will be presented in a forthcoming paper \citep{alam2018}. We will then use STIS, WFC3 and IRAC data to retrieve the composition and structure of this planet's atmosphere. Additional constraints will be placed by \textit{JWST}, which will observe WASP-52b in emission spectroscopy as part of the GTO program 1224 (PI Birkmann). 



\acknowledgments

\small{Based on observations made with the NASA/ESA Hubble Space Telescope, obtained from the data archive at the Space Telescope Science Institute. STScI is operated by the Association of Universities for Research in Astronomy, Inc. under NASA contract NAS 5-26555. These observations are associated with program GO 14260. The authors thank Neale Gibson and Dan Foreman-Mackey for their helpful suggestions about the use of \texttt{emcee} and of Gaussian processes for transit modeling, and Patricio Cubillos for his help on specific MCMC issues.}



Facilities: \facility{{\em HST}/WFC3}.



\bibliography{biblio}

\end{document}